\documentclass[review]{elsarticle}
\usepackage{lineno,hyperref}
\modulolinenumbers[5]
\usepackage{amsmath}
\usepackage{amsthm}
\usepackage{dsfont}
\usepackage{amssymb}
\usepackage{graphicx,color}
\usepackage{extarrows}
\usepackage{mathrsfs}
\usepackage{booktabs}
\usepackage{multirow}
\usepackage[justification=raggedright]{caption}
\usepackage{threeparttable}
\usepackage[misc]{ifsym}
\usepackage[linesnumbered,ruled]{algorithm2e}
\usepackage{array}
\usepackage{graphicx}
\usepackage{subfig}
\usepackage{setspace}
\usepackage{float}

\makeatletter
\@addtoreset{equation}{section}
\makeatother

\biboptions{numbers,sort&compress}

\begin{document}
	\begin{frontmatter}
		
		\title{\bf\Large Solving Fokker-Planck-Kolmogorov Equation by Distribution Self-adaptation Normalized Physics-informed Neural Networks}
		
		\author[addr1]{Yi Zhang}
		\author[addr2]{Yiting Duan}
		\author[addr1]{Xiangjun Wang}
		\author[addr1,addr3]{Zhikun Zhang\corref{cor1}}
		
		\ead{zhikunzhang@nwpu.edu.cn}
		\cortext[cor1]{Corresponding author}
		
		\address[addr1]{School of Mathematics and Statistics,\\
			Huazhong University of Science and Technology, Wuhan 430074, China}
		\address[addr2]{Translational Health Research Institute,\\ Western Sydney University, Sydney, NSW 2751, Australia}
		\address[addr3]{School of Mathematics and Statistics,\\
			Northwestern Polytechnical University, Xi’an 710072, China}
		
		\begin{abstract}
			Stochastic dynamical systems provide essential mathematical frameworks for modeling complex real-world phenomena. The Fokker–Planck–Kolmogorov (FPK) equation governs the evolution of probability density functions associated with stochastic system trajectories. Developing robust numerical methods for solving the FPK equation is critical for understanding and predicting stochastic behavior. Here, we introduce the distribution self-adaptive normalized physics-informed neural network (DSN-PINNs) for solving time-dependent FPK equations through the integration of soft normalization constraints with adaptive resampling strategies. Specifically, we employ a normalization-enhanced PINN model in a pretraining phase to establish the solution's global structure and scale, generating a reliable prior distribution. Subsequently, guided by this prior, we dynamically reallocate training points via weighted kernel density estimation, concentrating computational resources on regions most representative of the underlying probability distribution throughout the learning process. The key innovation lies in our method's ability to exploit the intrinsic structural properties of stochastic dynamics while maintaining computational accuracy and implementation simplicity. We demonstrate the framework's effectiveness through comprehensive numerical experiments and comparative analyses with existing methods, including validation on real-world economic datasets.
		\end{abstract}
		
		
	\end{frontmatter}
	
	\section{Introduction}\label{sec1}
	Deterministic characterization of stochastic dynamics through quantities that encapsulate dynamical information is fundamental for understanding the stochastic differential equation (SDE). Among these quantities, the probability density function (PDF) of solution trajectories plays a pivotal role. The Fokker-Planck-Kolmogorov (FPK) equation, a second-order parabolic partial differential equation (PDE), provides a comprehensive framework for describing the temporal evolution of these PDFs, thereby enabling systematic analysis of complex stochastic systems. This mathematical formalism has found extensive applications across diverse scientific and engineering domains, including laser light statistics\cite{Risken1996}, system control and simulation\cite{Ismail2023}, and chemical reaction networks\cite{Lunz2021}, among others. However, obtaining analytical solutions to FPK equations remains a formidable challenge, as closed-form expressions are rarely attainable. This limitation has motivated the development of numerous numerical approaches for approximating FPK solutions\cite{Pichler2013, Naprstek2014, Sepehrian2015}.
	
	Traditional numerical methods for FPK equations face dual challenges: the absence of well-defined boundary conditions and the curse of dimensionality that renders computations prohibitively expensive as system dimensions increase\cite{Jiang2015}. Recent advances in machine learning and computational capabilities have positioned neural network-based approaches as promising alternatives for solving PDEs, including FPK equations. Sirignano and Spiliopoulos pioneered the deep Galerkin method, employing deep neural networks to approximate solutions of parabolic PDEs\cite{Sirignano2018}. E and Yu introduced the deep Ritz method, which reformulates PDE problems within a variational framework\cite{Yu2018}. Guan developed a logistic basis function neural network with an iterative Gaussian neuron selection strategy for solving FPK equations under both Gaussian and non-Gaussian excitations\cite{Guan2023}. Li formulated a constrained optimization framework utilizing Monte Carlo simulation data to create a data-driven FPK solver\cite{Li2018}, which Zhai subsequently extended to a mesh-free implementation\cite{Zhai2022}.
	
	Physics-informed neural networks (PINNs), pioneered by Raissi, have recently emerged as powerful solvers for PDEs by embedding physical constraints directly into the neural network loss function\cite{Raissi2019}. Unlike traditional numerical methods, PINNs offer the distinct advantage of generating continuous solutions across the entire spatiotemporal domain without requiring computationally expensive mesh construction. Despite these advantages, early implementations of PINNs encountered significant challenges when applied to the FPK equation, particularly the tendency to converge to trivial solutions. Xu observed that standard PINNs solving FPK equations, which incorporate only the PDE and boundary conditions into the loss function, might frequently converge to the zero solution\cite{Xu2020}. To address this limitation, they introduced a normalization constraint to enforce integral preservation and incorporated penalty terms to steer the optimization away from local minima. Building on this foundation, Li and Meredith identified that while neural networks can effectively capture the shape of FPK solutions, they often struggle to learn the appropriate scale, primarily due to insufficient temporal regularization from the first-order derivative term\cite{Li2023}. Their solution involved augmenting the training process with Monte Carlo-sampled anchor points to better guide the learning of distribution magnitudes. More recently, Peng\cite{peng2024} developed a dual-network architecture that leverages prior knowledge from Monte Carlo simulations during a pretraining phase, significantly enhancing both training stability and solution accuracy. Complementary advances have emerged in high-dimensional sampling techniques\cite{Tang2022, Xiao2024, Liang2024}, further expanding the applicability of PINNs to complex systems.
	
	From an inverse modeling perspective, Chen explored the problem of inferring FPK dynamics from observed particle trajectories\cite{Chen2021}, successfully employing Kullback-Leibler divergence as a training objective to reconstruct governing stochastic dynamics from sparse empirical observations. Additionally, to address computational challenges in high-dimensional FPK problems, Khodabakhsh and Pourtakdoust introduced a dimension-reduced formulation of the FPK equation, enabling more scalable computations for practical applications\cite{Khodabakhsh2024}. Collectively, these advances demonstrate the rapid evolution of PINNs-based approaches for FPK equations, transforming them from methods prone to trivial solutions into robust computational tools capable of handling complex, high-dimensional stochastic systems.
	
	Training PINNs might fail when applied to strongly nonlinear or higher-order time-dependent PDEs, which often exhibit sharp transitions in the physical domain and pose significant challenges for accurate modeling\cite{Krishnapriyan2021, Basir2023, Rathore2024}. To address these limitations, existing approaches fall into three main categories: adaptive sampling strategies, modified loss functions, and improved network architectures. Adaptive sampling or resampling methods dynamically adjust training points to better capture complex behaviors and mitigate PINN failure modes\cite{Daw2023, Zhao2020}. Loss functions can be tailored for specific scenarios—for instance, self-adaptive PINNs treat loss weights as trainable parameters to emphasize regions with sharp features\cite{Mcclenny2023}. Architectural improvements include differentiable adversarial self-adaptive PINNs\cite{ZhangGT2023} and adaptive activation functions\cite{Jagtap2020}, both of which enhance model performance. Building on these advances, we propose a method that combines adaptive sampling with loss-balancing mechanisms specifically tailored to time-dependent FPK equations, aiming to achieve more accurate and effective solutions.
	
	In this study, we develop the distribution self-adaptive normalized physics-informed neural networks (DSN-PINNs) to solve time-dependent FPK equations. Recognizing the critical role of normalization conditions in ensuring solution accuracy, we introduce a normalization design that imposes soft constraints on the loss function, offering both computational accuracy and ease of implementation. These normalization-enhanced PINNs are employed during a pretraining phase, enabling the network to initially capture both the shape and scale of FPK solutions. The approximate distribution obtained from this phase then serves as a prior to guide subsequent resampling procedures. Unlike approaches that employ generative models for adaptive training set updates\cite{Feng2022} or rely on Monte Carlo simulations as pretraining data\cite{peng2024}, our method leverages stochastic dynamics information directly, eliminating the need for Monte Carlo simulations during training. This distinction is crucial: Monte Carlo simulation requires storing numerous sample trajectories, imposing a substantial computational burden. In contrast, our first-stage approach provides a prior compared to traditional simulation-based methods. The second stage involves strategically resampling training points based on the estimated probability distributions. For any fixed time, we employ weighted kernel density estimation to approximate the marginal distribution and reallocate training points accordingly, assigning higher sampling probabilities to regions where the target distribution function exhibits larger values. To prevent excessive clustering of points, we implement a mixture strategy combined with an iterative algorithm that progressively refines solution accuracy within our framework.	
	
	The remainder of this paper is organized as follows. Section~\ref{sec2} presents the necessary preliminaries and mathematical foundations. Section~\ref{sec3} describes our proposed approach, including the distribution self-adaptive normalized strategy, the normalization design, and the development of the DSN-PINNs. Section~\ref{sec4} outlines the numerical experiment settings and implementation details. Section~\ref{sec5} provides comparisons and analyses of the experimental results. Section~\ref{sec6} reports additional experiments on real-world datasets to further validate the proposed method. Finally, Section~\ref{sec7} concludes the paper and discusses potential directions for future research.

	\section{Preliminaries}\label{sec2}
	In this section, we present the fundamental concepts and background that will be used throughout this study.
	
	\subsection{The Fokker-Planck-Kolmogorov Equation}
	Consider an It\^{o} process $\{x_t\}_{t\in T}\in \mathbb{R}^n$ defined on some probability space $(\Omega, \mathcal{F}, \mathbb{P})$ describing the state of a stochastic dynamic system in $\mathbb{R}^n$. It satisfies the following equation
	\begin{equation}\label{SDE}
		\dot{\mathbf{x}}_t = \mathbf{b}(\mathbf{x}_t, t) + \boldsymbol{\sigma}(\mathbf{x}_t, t)\dot{\mathbf{B}}_t(\omega), \quad t \geq 0,
	\end{equation}
	where $\mathbf{b}: \mathbb{R}^n \rightarrow \mathbb{R}^n$ represents an n-dimensional drift vector function and $\boldsymbol{\sigma}: \mathbb{R}^n \rightarrow \mathbb{R}^{n\times m}$ is known as the diffusion matrix function. $\dot{\mathbf{B}}_t(\omega)$ is an m-dimensional Gaussian white noise vector, characterized by the spectral density function, i.e., the Fourier transform $\mathbb{F}$ of its covariance function $\mathbb{E}[\dot{\mathbf{B}}_t\dot{\mathbf{B}}_s]$, for $\dot{\mathbf{B}}_t$ that has a constant absolute value. 
	
	The generator $\mathcal{L}$ of the system \ref{SDE} is defined on the Hilbert space $L^2(\mathbb{R})$, as
	\begin{align}
		\mathcal{L}u &= \mathbf{b} \cdot (\nabla u) + \frac{1}{2} \operatorname{tr}[\boldsymbol{\sigma} \boldsymbol{\sigma}^T \mathbf{G}(u)] \\
		&= \sum_i b_i \frac{\partial u}{\partial x_i} + \frac{1}{2} \sum_{i,j} (\sigma \sigma^T)_{i,j} \frac{\partial^2 u}{\partial x_i \partial x_j}, \quad u \in H_0^2(\mathbb{R}^n),
	\end{align}
	which is assumed to satisfy the uniform ellipticity condition. The elements of the matrix $\boldsymbol{\sigma} \boldsymbol{\sigma}^T$ are given by $(\sigma \sigma^T)_{i,j} = \sum\limits_{k=1}^n \sigma_{ik} \sigma_{kj}$. Then the FPK equation describing the time evolving PDF of $\{X_t\}$ can be written as
	\begin{align}\label{FPK}
		\partial_t p(\mathbf{x}, t) &= \mathcal{L}^* p(\mathbf{x}, t) = - \sum\limits_{i=1}^n \frac{\partial}{\partial x_i}(b_i p)+\frac{1}{2}\sum\limits_{i,j=1}^n \frac{\partial}{\partial x_i \partial x_j} \left[ (\sigma \sigma^T)_{i,j} p \right],
	\end{align}
	where the initial condition $p(\mathbf{x}, t_0) = p_0(\mathbf{x})$ can be deterministically modeled by Dirac's delta distribution or described with uncertainty through a probability distribution. The operator $\mathcal{L}^*$ is the adjoint of $\mathcal{L}$, satisfying the inner product relationship $(\mathcal{L}f, g)_{L^2} = (f, \mathcal{L}^*g)_{L^2}$. Equation \ref{FPK} represents the time-dependent FPK equation defined in an unbounded domain, with the boundary condition $\lim\limits_{x\rightarrow \infty} p(\mathbf{x}) = 0$ and the normalization condition $\int_{\mathbb{R}^n} p(\mathbf{x}, t)\mathrm{d} \mathbf{x} \equiv 1$, where $p(\mathbf{x}, t) \geq 0$. If $\mathcal{L}^* p = 0$, the equation becomes the steady-state FPK equation, and the stationary solution (if it exists) implies the asymptotic behavior of the system \ref{SDE} as $t \to \infty$. The above time-dependent FPK equation can be summarized as
	\begin{equation}\label{FPK_summary}
		\left\{
		\begin{array}{l}
			p_t - \mathcal{L}^* p = 0, \quad \mathbf{x} \in \mathcal{X}\subset\mathbb{R}^n , \ t \in [0, T],\\[8pt]
			p(\mathbf{x}, t) \rightarrow 0 \ \text{as} \ \|\mathbf{x}\| \to \infty, \\[8pt]
			p(\mathbf{x}, t_0) = p_0(\mathbf{x}), \quad \mathbf{x} \in \mathcal{X}, \\[8pt]
			\int_{\mathcal{Q} } p_t(\mathbf{x}) \, \mathrm{d} \boldsymbol{x} = 1, \quad p(\mathbf{x}, t) \geq 0, \ \mathbf{x} \in \mathcal{X} , \ t \in [0, T].
		\end{array}
		\right.
	\end{equation}
	In practice, one can assume Dirichlet zero boundary conditions over a domain $\mathcal{X}$ large enough to cover all high-density areas with sufficient margin.
	
	\subsection{Physics-informed neural networks}
	Consider the initial-boundary value problem \ref{FPK_summary}, rewritten with the FPK differential operator $\mathcal{N}(p(\boldsymbol{x}, t)) = 0$ and the boundary operator $\mathcal{B}(p(\boldsymbol{x}, t)) = g(\mathbf{x}, t)$. Let $p(\mathbf{x}, t): \bar{\mathcal{Q}} \to \mathbb{R}$ be the desired solution, which is approximated by the output $\hat{p}(\mathbf{x}, t; \mathbf{w})$ of a deep neural network with inputs $\mathbf{x}$ and $t$. The network weights $\mathbf{w}$ can be tuned by minimizing the general loss function of type \ref{loss_general} that penalizes the output for not adhering to the physical priors \ref{FPK_summary} through standard gradient descent procedures as
	\begin{align}\label{loss_general} 
		\mathcal{L}oss &= \mathcal{L}oss_r + \mathcal{L}oss_b +\mathcal{L}oss_0,
	\end{align}
	where
	\begin{equation}
		\left\{
		\begin{aligned}
			\mathcal{L}oss_r &= \frac{1}{N_r}\sum\limits_{i=1}^{N_r}\left|\mathcal{N}(\hat{p}(\mathbf{x}_r^i,t_r^i; \mathbf{w}))\right|^2, \\[8pt]
			\mathcal{L}oss_b &= \frac{1}{N_b}\sum\limits_{i=1}^{N_b}\left|\mathcal{B}(\hat{p}(\mathbf{x}_b^i,t_b^i; \mathbf{w}))-g(\mathbf{x}_b^i, t_b^i)\right|^2,  \\[8pt]
			\mathcal{L}oss_0 &= \frac{1}{N_0}\sum\limits_{i=1}^{N_0}\left|\hat{p}(\mathbf{x}_0^i,0;\mathbf{w}))-p_0(\mathbf{x}_0^i)\right|^2,
		\end{aligned}
		\right.
	\end{equation}
	where $\{\mathbf{x}_i^r\}, \{\mathbf{x}_i^b\}, \{\mathbf{x}_i^0\}$ represent the PDE residue points, boundary points, and initial points within the selected regions $\mathcal{Q}$, respectively. Traditional PINNs treat these three types of points as training inputs to a fully connected neural network and train the network according to the loss function \ref{loss_general}.
	
	\subsection{Euler–Maruyama Methods}\label{MC_EM}
	Most stochastic dynamics or differential equations lack analytic solutions and therefore require Monte Carlo simulations to generate large enough sample paths and obtain the distribution at different time steps\cite{Kloeden2011}. Euler–Maruyama is a direct method for approximating the numerical solution of \ref{SDE}. For simplicity, the following equations represent the scalar version of the approximation without vector notation. The Euler–Maruyama approximation to the analytic solution ${X_t}$ is a Markov chain ${Y_t}$ defined as
	\begin{align}
		Y_{i+1} &= Y_i + b(Y_i, t_i,)(t_{i+1}-t_i)+\sigma(y_i, t_i)(B_{i+1}-B_i)\\
		& = Y_i + b(y_i, t_i,)(t_{i+1}-t_i)+ \sqrt{t_{i+1}-t_i}Z_i,\\
		Z_i &\sim \mathcal{N}(0, 1),\ i=0,1,..., N, 
	\end{align}
	where $t_0\leq t_1 \leq \dots \leq t_N = T, \ \Delta t = t_{i+1}- t_i = T/N$. The Milstein method is a more accurate scheme since the Euler method is actually set by
	\begin{equation}
		\int_{t}^{t+\Delta t}\sigma(X_u, u)\mathrm{d}B_u = \sigma(X_t, t)\int_{t}^{t+\Delta t}\mathrm{d}B_u,
	\end{equation}
	while Milstein improves the estimation accuracy by incorporating second-order terms using It\^{o}'s lemma
	\begin{equation}
		Y_{i+1} = Y_i + b(Y_i, t_i,)\Delta t+\sigma(y_i, t_i)(B_{i+1}-B_i)+ \frac{1}{2}\sigma\sigma_x[((B_{i+1}-B_i)^2-\Delta t],
	\end{equation}
	where $\sigma \in C^2(\mathbb{R}^d)$.
	
	\section{Methodology}\label{sec3}
	
	Our method firstly extracts the fundamental distribution information of the stochastic dynamics via PINNs with normalization design, then implements a resampling scheme based on the distribution in the previous stage. 
	
	\subsection{Normalization Design}
	As previously discussed, due to the problem in accurately capturing the correct scale and mitigating the trivial solution, we propose a modified normalization design that imposes the integral constraint as a soft regularization term in the loss function. In prior methods, the normalization condition was viewed as a regularization term, reflecting the fact that the solution $p_t(x)$ is a probability density. Specifically, this is achieved by adding the term,
	\begin{equation}\label{raw_normlization}
		\mathcal{L}oss_n = \left| \sum_{r=1}^n \Delta x_r^i \cdot \hat{p}(x_r^i,t_r^i;\mathbf{w}) - 1 \right|^2,
	\end{equation} 
	to the loss function. However, \ref{raw_normlization} requires constructing a discrete grid over the entire domain and performing a summation over all grid points, leading to an exponential growth in computational cost as the dimensionality increases. Hence, we use a normalization term \ref{normaliza} approximated by an average computed only over the existing sample points. This way, the computational complexity scales linearly with the number of samples, significantly reducing the overall cost as
	\begin{equation}\label{normaliza}
		\mathcal{L}oss_{n} =  \left|\bar{p}(\mathbf{x},t;\mathbf{w})-\frac{C_T}{\mathcal{M}(\mathcal{Q})}\right|^2,
	\end{equation}
	where $C$ is a positive constant and  $\mathcal{M}(\mathcal{Q})$ denotes the Lebesgue measure of the domain $\mathcal{Q}$. The non-zero probability solution $p(x,t)$ is generally concentrated in an unknown region $\mathcal{Q}\subset \mathbb{R}^n\times[0, T]$, and the scale of this region can be estimated through a single rough Monte Carlo simulation. The term $\bar{p}(\mathbf{x}, t; \mathbf{w})$ represents the average value of $\hat{p}(\mathbf{x}, t; \mathbf{w})$ over the domain $\mathcal{Q}$ and can be approximated by
	\begin{equation}\label{approx_avg}
		\bar{p}(\mathbf{x}, t; \mathbf{w}) = \lim_{N_r, N_b, N_0 \to \infty} \frac{1}{N_r + N_b + N_0} \left[ \sum_{i=1}^{N_r} \hat{p}(\mathbf{x}_r^i, t_r^i; \mathbf{w}) + \sum_{i=1}^{N_b} \hat{p}(\mathbf{x}_b^i, t_b^i; \mathbf{w}) + \sum_{i=1}^{N_0} \hat{p}(\mathbf{x}_0^i, t_0^i; \mathbf{w}) \right].
	\end{equation}
	
	The normalization design guarantees that the network outputs are properly normalized. If $p^*(\mathbf{x}, t; \mathbf{w})$ denotes the outputs satisfying the normalization condition, we define the following transformation to enforce normalization
	\begin{align}\label{hard}
		p^*(\mathbf{x}, t; \mathbf{w}) &= \hat{p}(\mathbf{x}, t; \mathbf{w}) - \frac{\int_{\mathcal{Q}} \hat{p}(\mathbf{x}, t; \mathbf{w}) \, \mathrm{d}\mathbf{x} \, \mathrm{d}t - C_T}{\mathcal{M}(\mathcal{Q})} \\ 
		&= \hat{p}(\mathbf{x}, t; \mathbf{w}) - \bar{p}(\mathbf{x}, t; \mathbf{w}) + \frac{C_T}{\mathcal{M}(\mathcal{Q})}.
	\end{align}
	Integrating both sides over $\mathcal{Q}$ yields
	\begin{equation}
		\int_{\mathcal{Q}} p^*(\mathbf{x}, t; \mathbf{w}) \, \mathrm{d}\mathbf{x} \, \mathrm{d}t 
		= \int_{\mathcal{Q}} \hat{p}(\mathbf{x}, t; \mathbf{w}) \, \mathrm{d}\mathbf{x} \, \mathrm{d}t 
		- \mathcal{M}(\mathcal{Q}) \cdot \frac{\int_{\mathcal{Q}} \hat{p}(\mathbf{x}, t; \mathbf{w}) \, \mathrm{d}\mathbf{x} \, \mathrm{d}t - C_T}{\mathcal{M}(\mathcal{Q})} 
		= C_T.
	\end{equation}
	The deviation between the raw and normalized network outputs is quantified by
	\begin{equation}
		\left|p^*(\mathbf{x}, t; \mathbf{w})-\hat{p}(\mathbf{x}, t; \mathbf{w})\right| = \left|\bar{p}(\mathbf{x},t;\mathbf{w})-\frac{C_T}{\mathcal{M}(\mathcal{Q})}\right|.
	\end{equation}
	This deviation is incorporated into the loss function to penalize violations of the normalization constraint. Additionally, the constant $C$ corresponds to the temporal span of the system evolution. Since $\forall t \in [0, T]$, we have 
	\begin{align}
		\int_{\mathbb{R}^n} p(\mathbf{x}, t) \, \mathrm{d}\mathbf{x} &= 1,
	\end{align}
	and then obtained
	\begin{align}   
		\int_{0}^{T} \int_{\mathbb{R}^n} p(\mathbf{x}, t) \, \mathrm{d}\mathbf{x} \, \mathrm{d}t &= C_T.
	\end{align}
	
	Compared to \ref{raw_normlization} and the anchor method mentioned previously, our normalization design largely reduces computational cost and is easier to implement since neither a mesh grid nor expensive Monte Carlo methods are required in the training stage. Nevertheless, the estimate in \ref{approx_avg} becomes inaccurate when the sample points are concentrated in a subregion. We refer to the sample points obtained by non-adaptive methods, i.e., uniform sampling, within $\mathcal{Q}$ as the base points $\mathcal{D}_b$. Therefore, we employ base points to train the PINNs with the normalization design PINNs (N-PINNs) in a pretraining (warm start) phase to initially extract the distribution information. This distribution then serves as a prior in the subsequent distribution self-adaptation PINNs (D-PINNs) resampling procedure, guiding the selection of new sample points through the global characteristics of the target density.
	
	\begin{figure}[t]
		\centering
		\includegraphics[width=0.85\textwidth]{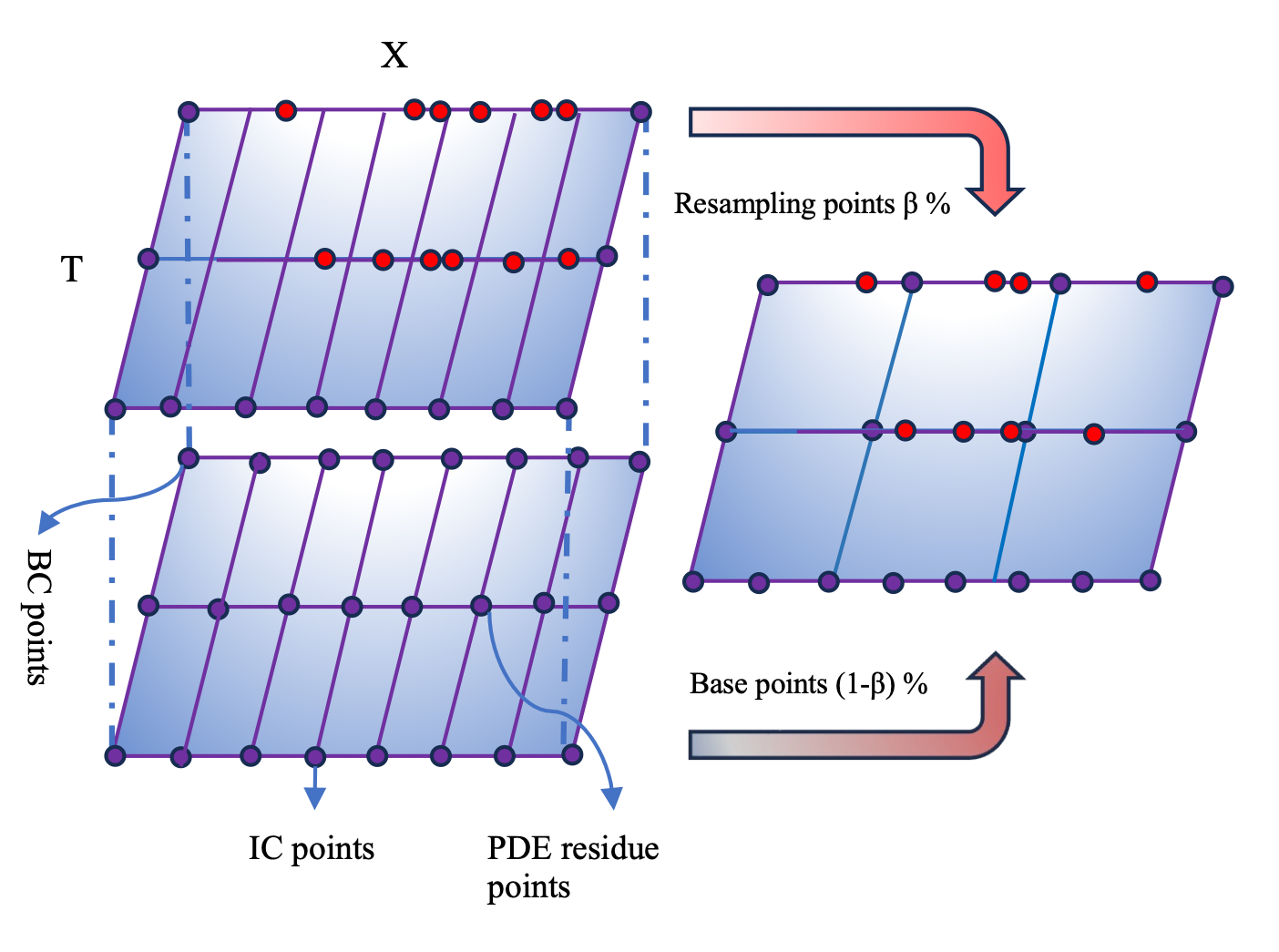}  
		\caption{Schematic Diagram of Mixture Strategy in DSN-PINNs.}
		\label{sample process}
	\end{figure}
	
	\subsection{Distribution Self-Adaptive Normalized PINNs}
	The selection of training points is crucial. Non-adaptive methods, such as uniform sampling, Latin Hypercube Sampling, and Sobol sequences, may not provide sufficient accuracy in complex problems. Base points obtained by these methods are particularly suitable for N-PINNs because the approximation in equation \ref{approx_avg1} is precisely a Monte Carlo estimator. Only when these base points are sufficiently dense and well-distributed does the estimator achieve low variance; otherwise, clustered or uneven sampling increases Monte Carlo error and undermines the normalization accuracy. The corresponding representation is
	\begin{equation}\label{approx_avg1}
		\int_{\mathcal{Q}} \hat{p}(\mathbf{x}, t; \mathbf{w}) \, \mathrm{d}\mathcal{Q} = \lim_{N_r, N_b, N_0 \to \infty} \frac{\mathcal{M}(\mathcal{Q})}{N_r + N_b + N_0} \left[ \sum_{i=1}^{N_r} \hat{p}(\mathbf{x}_r^i, t_r^i; \mathbf{w}) + \sum_{i=1}^{N_b} \hat{p}(\mathbf{x}_b^i, t_b^i; \mathbf{w}) + \sum_{i=1}^{N_0} \hat{p}(\mathbf{x}_0^i, t_0^i; \mathbf{w}) \right].
	\end{equation}
	
	To improve PINNs' accuracy, adaptive resampling of the collocation points during the training process can be beneficial. Our proposed D-PINNs sufficiently exploit the intrinsic distribution information of the stochastic dynamics to drive an iterative adaptive resampling scheme. The distribution attained in the former iteration guides the resampling of training points throughout subsequent iterations, ensuring the points remain representative of the system's true dynamics.
	
	Suppose we first utilize base points $\mathcal{D}_b$ to pretrain N-PINNs and get an approximated solution $\hat{p}(\mathbf{x}, t; \mathbf{w})$. For any fixed time $t$, we could collect a snapshot of some discrete observations $\{\mathbf{x}_i, u_i\}_{i=1}^n$ taking values in $\mathbb{R}^d$ from the marginal probability distribution $p_t(\mathbf{x};\mathbf{w})$ with $u_i=\hat{p}_t(\mathbf{x}_i;\mathbf{w})$. Then we use the weighted kernel density estimate with a Gaussian kernel $\mathcal{K}(\cdot): \ \mathbb{R}^d \rightarrow \mathbb{R}$ and Scott's rule\cite{Scott2015} for bandwidth selection to update $\hat{p}(\mathbf{x},t;\mathbf{w})$ as
	\begin{equation}
		\hat{p}(\mathbf{x},t;\mathbf{w}) = \frac{1}{nh^d}\sum\limits_{i=1}^{n}u_i\mathcal{K}(\frac{\mathbf{x}-\mathbf{x}_i}{h}), \quad
		h = \sigma n^{-1/(d+4)},
	\end{equation}
	where $\sigma$ denotes the standard deviation of $\{\mathbf{x}_i\}_{i=1}^n$ and $d$ is the dimension of the system. Subsequently, $\hat{p}(\mathbf{x},t;\mathbf{w})$ is employed as weights for resampling the training points. A sufficiently large set of $N$ uniformly distributed points $\{\mathbf{y}_i\}_{i=1}^N$ is generated within the spatial domain for each fixed time $t$, and the normalized weight for each $\mathbf{y}_j$ is computed as
	\begin{equation}\label{weights}
		\alpha(\mathbf{y}_j) = \hat{p}_t(\mathbf{y}_j;\mathbf{w}) \Big/ \sum\limits_{i=1}^N\hat{p}_t(\mathbf{y}_i; \mathbf{w}).
	\end{equation}
	
	The weights \ref{weights} reflect the probability of selecting each point $\mathbf{y}_i$. Therefore, $n$ independent realizations $i_1, i_2,...,i_n$ are drawn from the categorical distribution defined by these probabilities as
	\begin{equation}
		\mathbb{P}(i_k = j) = \alpha(\mathbf{y}_j), \quad j=1,2,\dots,N,\quad k=1,2,\dots,n.
	\end{equation}
	
	To ensure that the final sample comprises $n$ distinct indices, duplicates are removed and additional samples are drawn until $n$ unique indices are obtained. Then the resampled set of training points at time $t$ is given by $\{\mathbf{y}_{i_1},\cdots, \mathbf{y}_{i_n}\}$. By applying the above procedure to each discrete time point $t$, a set of resampling spatiotemporal points $\mathcal{D}_r$ is obtained. Moreover, due to the potential concentration of training points in a limited region, which leaves other areas underrepresented, we introduced a mixture strategy, see Fig.~\ref{sample process}, which integrates the resampling points $\mathcal{D}_r$ and base points $\mathcal{D}_b$ in proportions $\beta$ and $(1-\beta)$, respectively. Thereby it constructing the composite training set $\mathcal{D}^*$ for PINNs training.  
	
	To enhance the resampling procedure, an iterative framework is introduced. The mean PDE residual $\mathcal{R}$ is evaluated at each iteration. The model parameters and training set $\mathcal{D}^*$ are updated only if $\mathcal{R}$ decreases by a threshold $\epsilon$ compared to the previous iteration. The procedure terminates when $\mathcal{R}$ achieves $N_{adap}$ successful reductions or reaches a maximum iteration limit $N_{max}$, ensuring robustness against infinite loops. See the complete pseudocode in Algorithm~\ref{pseudcode} for a summary of the resampling scheme.

	\begin{algorithm}
		\caption{Resampling Scheme of DSN-PINNs.}\label{pseudcode}
		
		\KwData{Base points $\mathcal{D}_b = \{(\mathbf{x}_i, t)\}_{i=1}^n$ with $M$ discrete time points $t \in [t_0, T]$, where $\mathbf{x}_i \in [-L, L]^d$; resampling ratio $\beta \in (0,1)$; adaptive iterations $N_{adap}$; maximum resampling number $N_{max}$ in each iteration; large enough $N$; threshold $\epsilon$.}

		\KwResult{Resampling points $\mathcal{D}^*$ and best model parameter $\mathbf{w}^*$.}
		
		Set $\mathcal{R}_{prev} \gets +\infty$, $\mathcal{D}^* \gets \mathcal{D}_b, \ k \gets 0, s \gets 0$\ \;
		
		\While{$k < N_{max}$ \textbf{and} $s<N_{adap}$}
		{
			Train PINNs (normalization is applied only for $k=0$ and $a=0$) on $\mathcal{D}^*$ to obtain approximated density $\hat{p}(\mathbf{x}, t; \mathbf{w})$ and calculate mean PDE residual $\mathcal{R} =\frac{1}{n}\sum_{i=1}^n|\mathcal{R}_i|$. Save current model parameters $\mathbf{w}$\;
			
			\If{$\mathcal{R}<\mathcal{R}_{prev}-\epsilon$}
			{
				Update best model parameters $\mathbf{w}^* \gets \mathbf{w}$ and $\mathcal{R}_{prev}\gets\mathcal{R}$\;
				
				$s = s+1$\;
				
				\For{$t = t_0 : \frac{T - t_0}{M} : T$}
				{
					Generate $\text{snapshot} \{\mathbf{x}_i, u_i\}_{i=1}^n$ from $\hat{p}(\mathbf{x}, t; \mathbf{w^*})$ with $u_i = \hat{p}_t(\mathbf{x}_i; \mathbf{w^*})$\;
					
					$\bar{\mathbf{x}} \gets \frac{1}{n} \sum_{i=1}^n \mathbf{x}_i, \ \sigma \gets \frac{1}{n} \sqrt{\sum_{i=1}^{n} \|\mathbf{x}_i - \bar{\mathbf{x}}\|^2}, \ h \gets \sigma n^{-\frac{1}{d+4}}$\;
					
					$\hat{p}(\mathbf{x}, t; \mathbf{w}^*) \gets \frac{1}{nh^d}\sum\limits_{i=1}^n u_i \mathcal{K}\left( \frac{\mathbf{x} - \mathbf{x}_i}{h} \right)$\;
					
					Generate candidate points $\{\mathbf{y}_i\}_{i=1}^N$ uniformly in $[-L, L]^d$\;
					
					$\alpha(\mathbf{y}_i) \gets \frac{\hat{p}(\mathbf{y}_i, t; \mathbf{w})}{\sum_{j=1}^N \hat{p}(\mathbf{y}_j, t; \mathbf{w})}$\;
					
					Sample $n$ indepent indices $\{i_1, \dots, i_{n}\} \sim \text{Categorical}(\alpha(\mathbf{y}_1),\cdots, \alpha(\mathbf{y}_N))$\;
					
					$\mathcal{D}_r\gets\{(\mathbf{y}_{i_k}, t)\}_{k=1}^n$\;
				}
				
				$n_r \gets \lfloor \beta \cdot n \rfloor$, \quad $n_b \gets n - n_r$\;
				$\mathcal{D}^* \gets \mathcal{D}_r^{(n_r)} \cup \mathcal{D}_b^{(n_b)}$\;
			}
			\Else
			{
				Keep model parameters and $\mathcal{D}^*$ unchanged\;
			}
			
			$k \gets k + 1$\;
		}
		
		\KwRet{$\mathcal{D}^*$, $\mathbf{w}^*$.}
	\end{algorithm}

	Although a mixture strategy is employed in the resampling scheme, overfitting may still occur if the original distribution is highly uneven, so that the self-adaptive loss weights technique could be beneficial, as shown in the following form.
	\begin{equation}
		\left\{
		\begin{aligned}
			\mathcal{L}oss_r(\mathbf{w}, \boldsymbol{\lambda}_r) &= \frac{1}{N_r}\sum\limits_{i=1}^{N_r} f(\lambda_r^i) \left| \mathcal{N}(\hat{p}(\mathbf{x}_r^i,t_r^i; \mathbf{w})) \right|^2, \\[8pt]
			\mathcal{L}oss_b(\mathbf{w}, \boldsymbol{\lambda}_b) &= \frac{1}{N_b}\sum\limits_{i=1}^{N_b} f(\lambda_b^i) \left| \mathcal{B}(\hat{p}(\mathbf{x}_b^i,t_b^i; \mathbf{w})) \right|^2,  \\[8pt]
			\mathcal{L}oss_0(\mathbf{w}, \boldsymbol{\lambda}_0) &= \frac{1}{N_0}\sum\limits_{i=1}^{N_0} f(\lambda_0^i) \left| \hat{p}(\mathbf{x}_0^i,0; \mathbf{w}) - p_0(\mathbf{x}_0^i) \right|^2,
			\\[8pt]
			\mathcal{L}oss_{n}(\mathbf{w}, \mu) &= \mu\left|\bar{p}(\mathbf{x},t;\mathbf{w})-\frac{C_T}{\mathcal{M}(\mathcal{Q})}\right|^2,
		\end{aligned}
		\right.
	\end{equation}
	where the strictly increasing $f: [0,\infty)\rightarrow \mathbb{R}$ stands for the self-adaptation mask function of pointwise weights $\boldsymbol{\lambda}$, while $\mu$ is the regularization coefficient. This technique originates from self-adaptive PINNs (S-PINNs) without a distribution adaptive strategy, and when integrated into D-PINNs and N-PINNs, it forms our final method named DSN-PINNs.
	
	\section{Numerical Experiment}\label{sec4}
	
	In this section, four numerical tests implemented using PyTorch are presented to demonstrate the effectiveness of the proposed algorithm. All trainable parameters are initialized via Xavier initialization\cite{glorot2010}, the Adam optimizer\cite{adam2014} is used for optimization, and the hyperbolic tangent (Tanh) function is employed as the activation function. In cases where the reference solution is not available in closed form, numerical simulations described in Section~\ref{MC_EM} are used to obtain the solution, thereby allowing us to assess the accuracy of our method. Note that the iterative framework in Algorithm~\ref{pseudcode} introduces additional training points by a factor of $\min\{N_{adap}, N_a\}\times\beta$ at each resampling step, so that when conducting comparative experiments, the original PINNs are trained on an augmented set containing $\min\{N_{adap}, N_a\}\times\beta$ times as many points as the base points $\mathcal{D}_b$.

	\begin{figure}[t]
		\centering
		\includegraphics[width=0.73\linewidth]{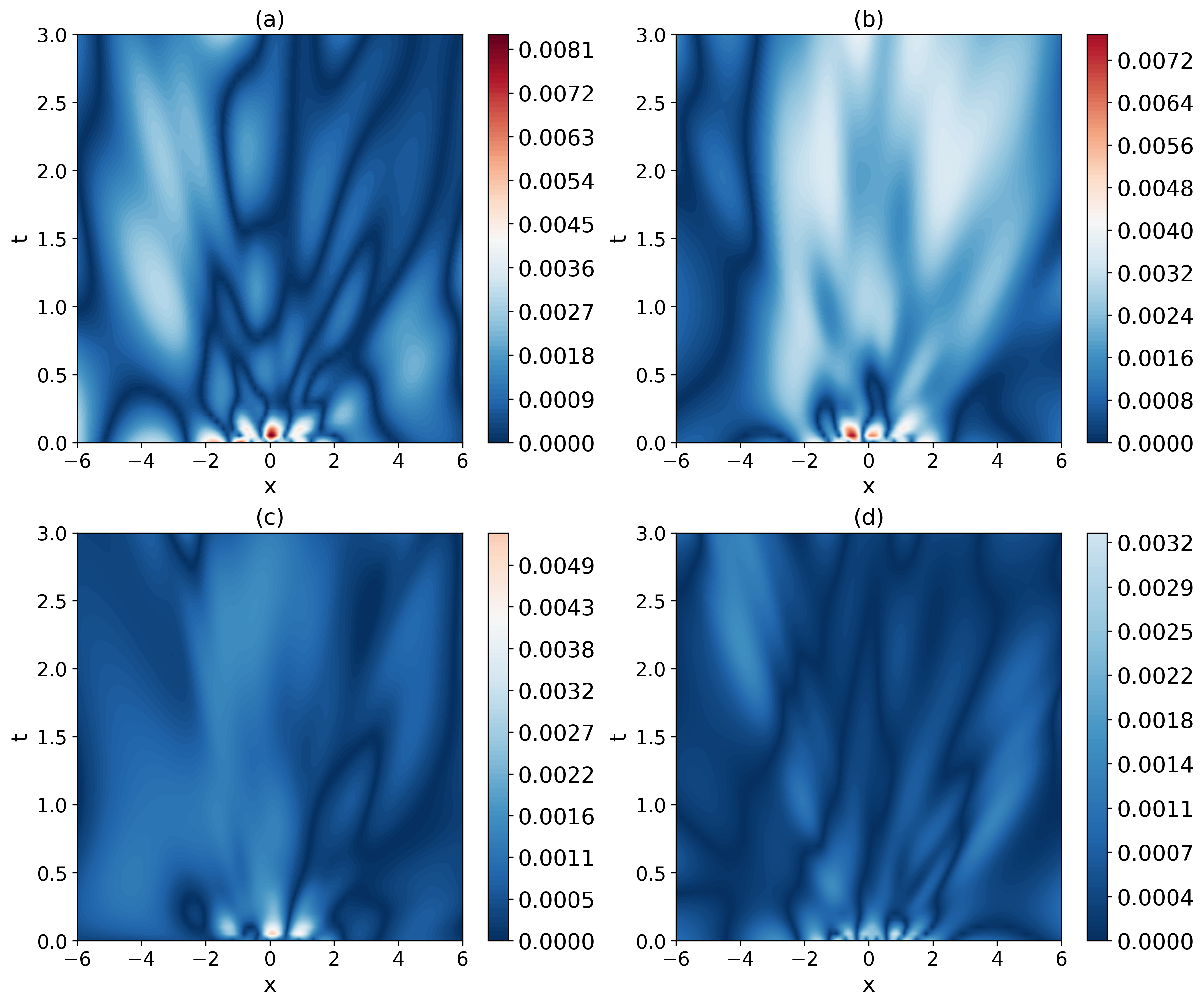}
		\caption{Absolute Errors for (a) Standard PINNs, (b) N-PINN, (c) D-PINNs, and (d) DSN-PINNs in Example 1. \label{LDBM_error}}
		\begin{minipage}{0.32\textwidth}
			\centering
			\includegraphics[width=\linewidth]{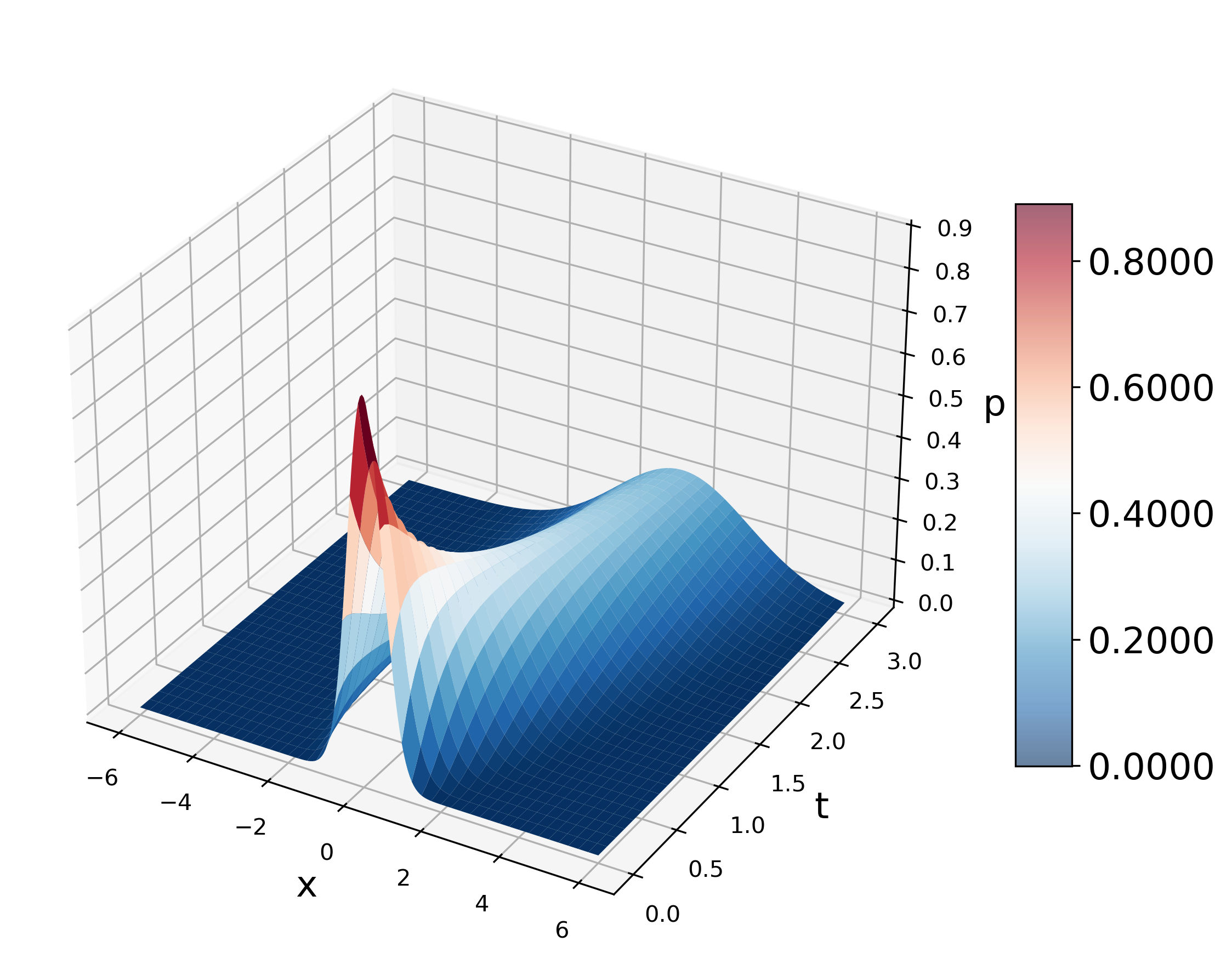}
		\end{minipage}  
		\hfill
		\begin{minipage}{0.32\textwidth}
			\centering
			\includegraphics[width=\linewidth]{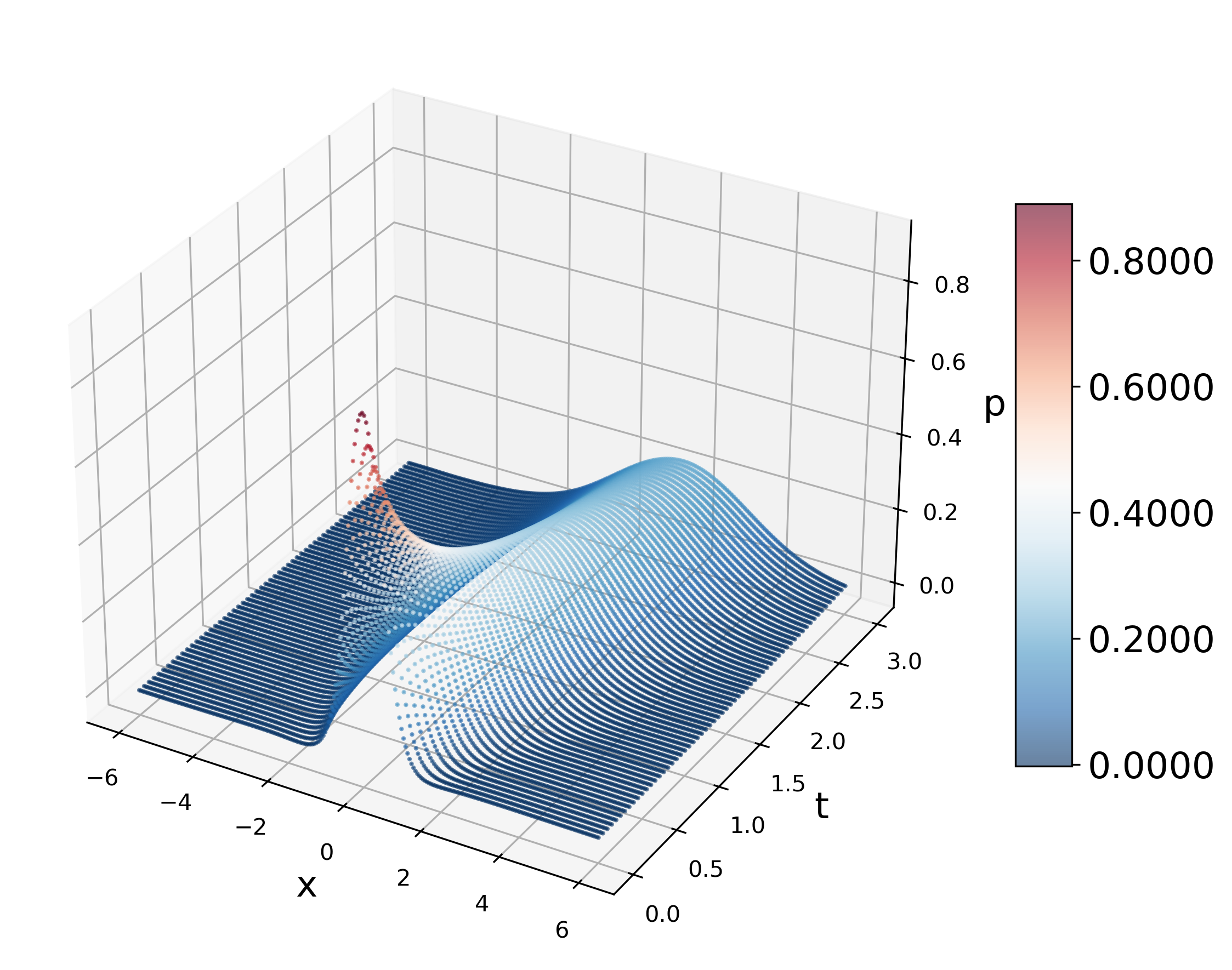}
		\end{minipage}
		\hfill
		\begin{minipage}{0.3\textwidth}
			\centering
			\includegraphics[width=\linewidth]{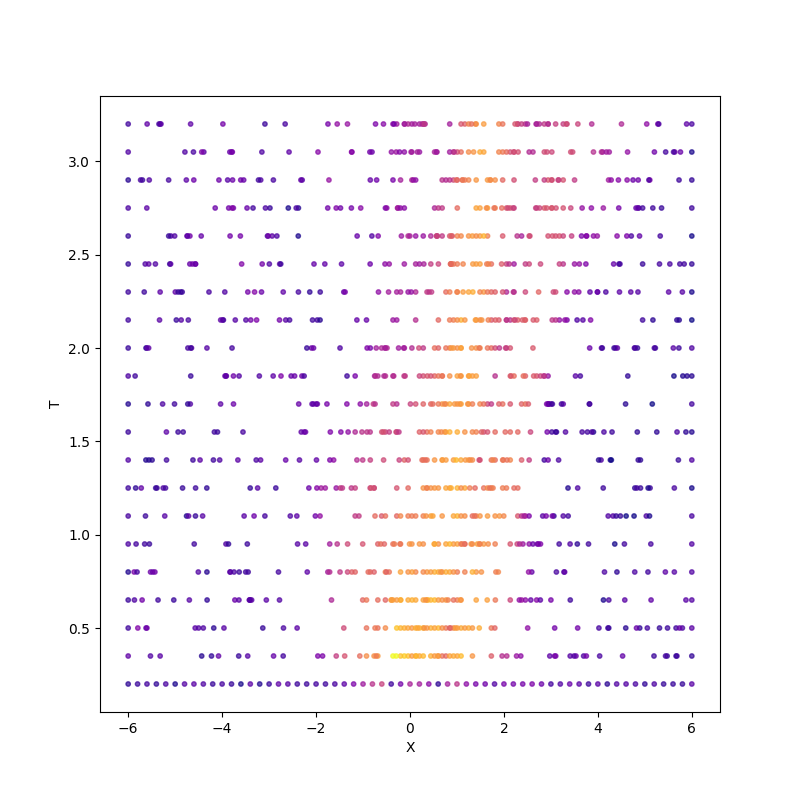}
		\end{minipage}  
		\caption{Left: Exact Solution of Example 1.\label{fig:exact_sol1} Middle: DSN-PINNs Prediction.
			\label{fig:LDBM_pred} Right: Resampling Points.\label{fig:resample_sa1}}
	\end{figure}

	\subsection{Example 1}
	As our first example, we start with a scaled Brownian motion with linear drift defined over the time interval $t\in [0, T]$ as
	\begin{align}\label{experiment1}
		\begin{cases}
			\mathrm{d} X_t = \mu \mathrm{d} t + \sigma \mathrm{d} B_t, \\[8pt]
			X_0 = 0.
		\end{cases}
	\end{align}
	The corresponding one-dimensional FPK equation of \ref{experiment1} is represented as
	\begin{equation}\label{FPK1}
		\frac{\partial p(x,t)}{\partial t} = -\mu \frac{\partial p(x,t)}{\partial x} + \frac{1}{2} \sigma^2 \frac{\partial^2 p(x,t)}{\partial x^2}.
	\end{equation}
	Generally, the initial condition of \ref{FPK1} is given as a deterministic delta function $\delta(x)$ or any probability distribution $p(x,0)\in \mathcal{L}^2(\mathcal{Q})$. A point distribution as the initial condition can cause difficulties during training, so we define the initial distribution as
	\begin{equation}
		p(x,0) = \frac{1}{\sqrt{0.4 \pi \times 0.2}} \exp \left( -\frac{(x - 0.1)^2}{0.4} \right).
	\end{equation}
	The exact solution yields
	\begin{equation}
		p(x,t) = \frac{1}{\sqrt{2 \pi (t+0.2)}} \exp \left( -\frac{(x -  \frac{t+0.2}{2})^2}{2(t+0.2)} \right),
	\end{equation}
	which is equivalent to shifting the time origin so that $t=0.2$ becomes the new starting point for the simulation.
	
	We set the spatiotemporal domain as $\mathcal{Q} = [-6,6] \times [0,3]$ with a resampling ratio $\beta = 0.5$. The neural network architecture consists of 3 hidden layers with 20 neurons each. The mask function $f(\cdot)$ in the self-adaptive loss weights is chosen as the Sigmoid function, and the learnable loss weights are initialized from the uniform distribution $\mathcal{U}[0,1]$. We construct the base training set $\mathcal{D}_b$ by uniformly sampling 160 initial, 80 boundary, and 50 spatial points at each of the 40 time steps, yielding $50 \times 40$ interior points. To better validate the proposed method, four models are compared: standard PINNs, N-PINNs, D-PINNs, and DSN-PINNs. In DSN-PINNs, the threshold $\epsilon = 5\times10^{-5}$, with at most $N_{\max}=5$ resampling numbers and $N_{\mathrm{adaptive}}=3$ adaptive iterations. Learning rates in the Adam optimizer are set to 0.001, 0.003, 0.005, and 0.005 for each model, respectively.
	
	Figure~\ref{LDBM_error} depicts the error heatmap between the numerical solution and the reference solution. Figure~\ref{fig:resample_sa1} illustrates the resampled points with their associated weights from the final iteration of DSN-PINNs. Brighter points correspond to higher weights, which tend to cluster in regions where the PDE is harder to approximate. This pattern aligns well with the exact solution and predictive solution also shown in Figure~\ref{fig:exact_sol1}, where complex dynamics lead to larger residuals and thus draw more training focus. The detailed results and comprehensive comparisons are presented in the following section.
	
	\subsection{Example 2}
	We next consider an example without a closed-form solution on the time interval $t\in [0, T]$ as
	\begin{align}\label{experiment2}
		\begin{cases}
			\mathrm{d} X_t = (X_t-X_t^3) \mathrm{d} t + \sigma(X_t) \mathrm{d} B_t, \\[8pt]
			X_0 = 0.
		\end{cases}
	\end{align}
	
	The corresponding one-dimensional FPK equation of \ref{experiment2} is
	\begin{equation}\label{FPK2}
		\frac{\partial p(x,t)}{\partial t} = -\frac{\partial}{\partial x} \left[ (x - x^3)p(x,t) \right] + \frac{1}{2} \frac{\partial^2}{\partial x^2} \left[ \sigma^2(x)p(x,t) \right].
	\end{equation}
	Let $\sigma(x) \equiv 1$. The initial distribution is given by a Gaussian with mean zero and variance $0.2$ as
	\begin{equation}
		p(x, 0) = \frac{1}{\sqrt{2\pi \times 0.2}} \exp\left( -\frac{x^2}{2\times 0.2} \right).
	\end{equation}
	
	\begin{figure}[t]
		\centering
		\includegraphics[width=0.7\linewidth]{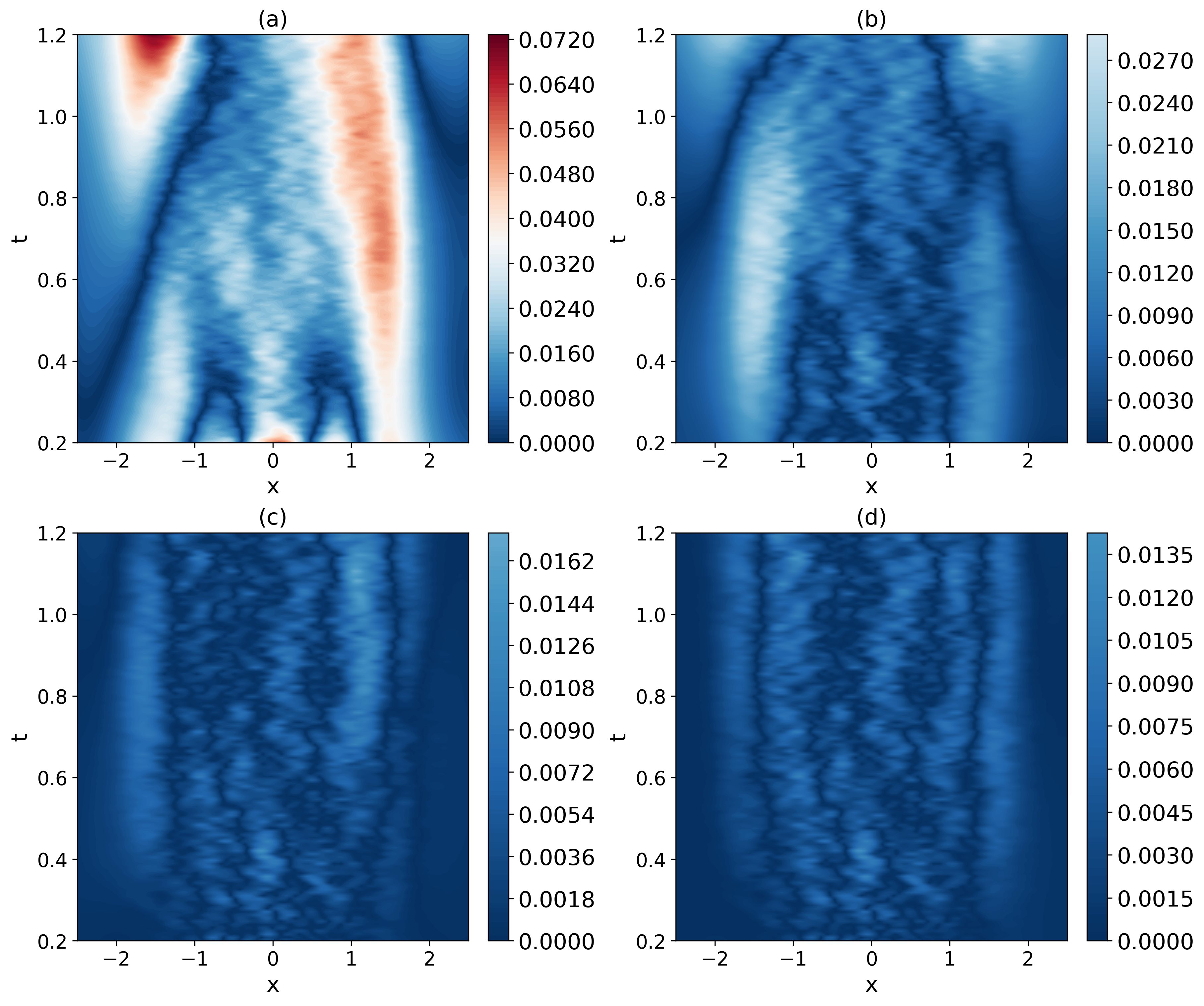}
		\caption{Absolute Errors for (a) Standard PINNs, (b) N-PINN, (c) D-PINNs, and (d) DSN-PINNs in Example 2. \label{Eg1_error}}
		\begin{minipage}{0.32\textwidth}
			\centering
			\includegraphics[width=\linewidth]{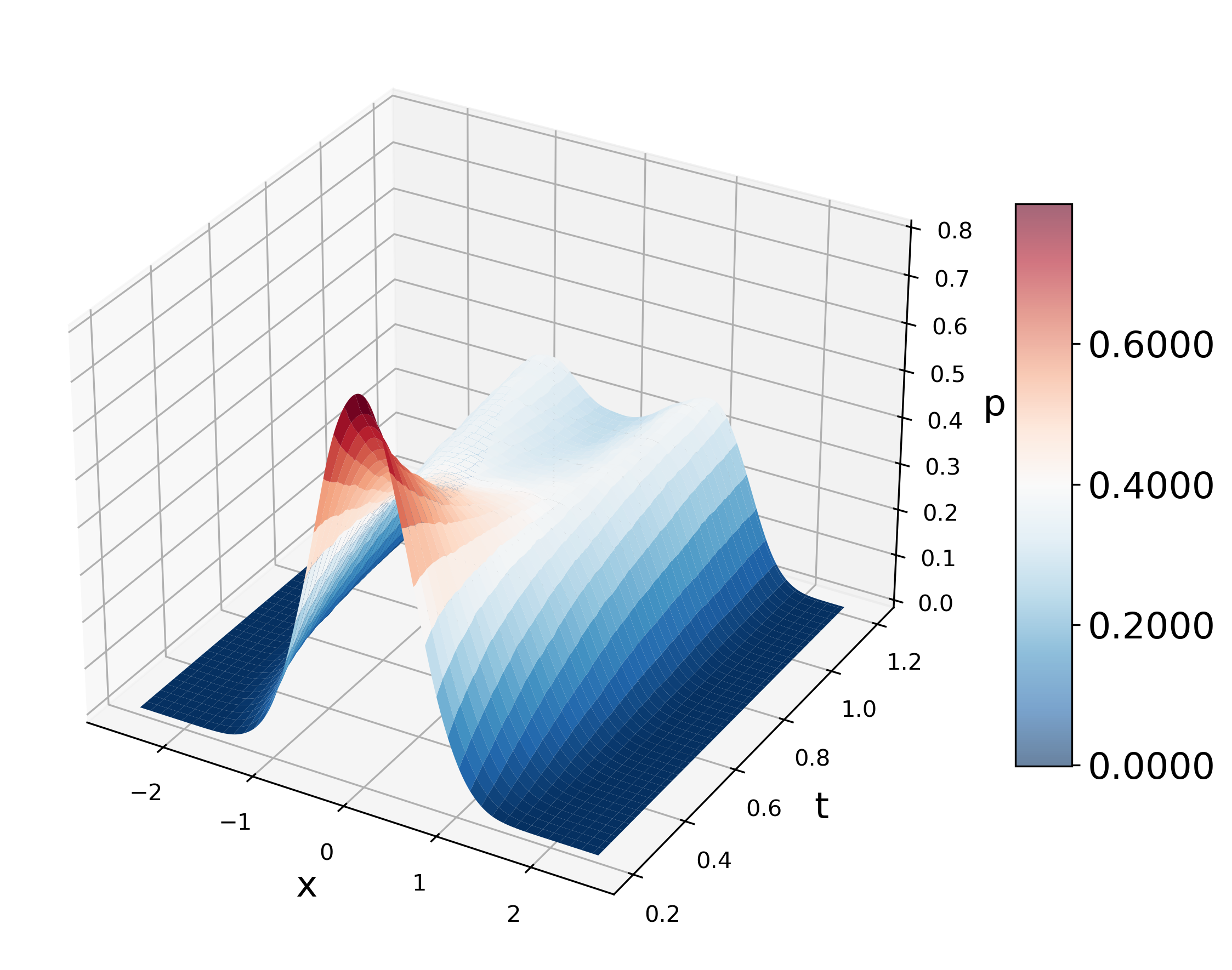}
		\end{minipage} 
		\hfill
		\begin{minipage}{0.32\textwidth}
			\centering
			\includegraphics[width=\linewidth]{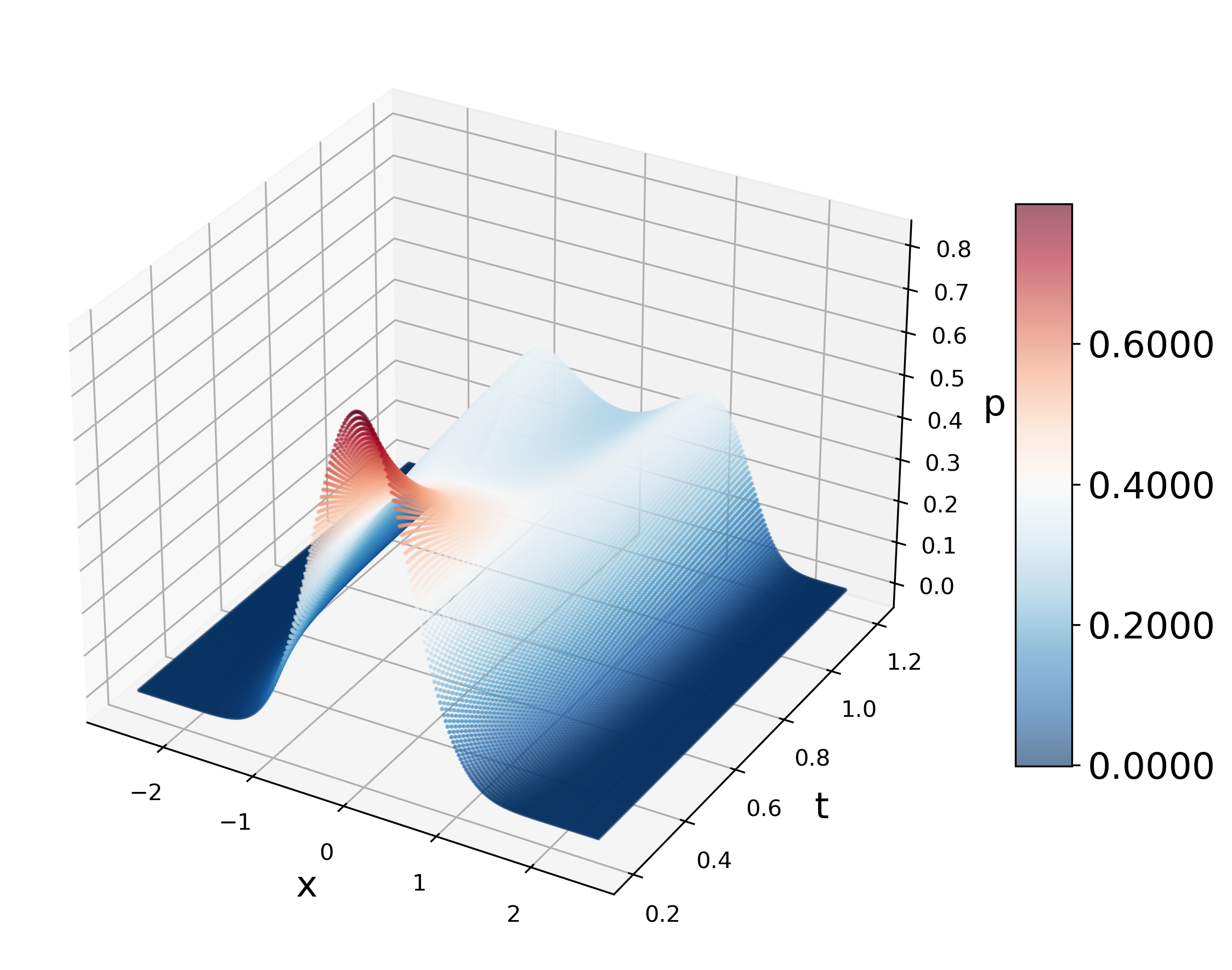}
		\end{minipage}
		\hfill
		\begin{minipage}{0.3\textwidth}
			\centering
			\includegraphics[width=\linewidth]{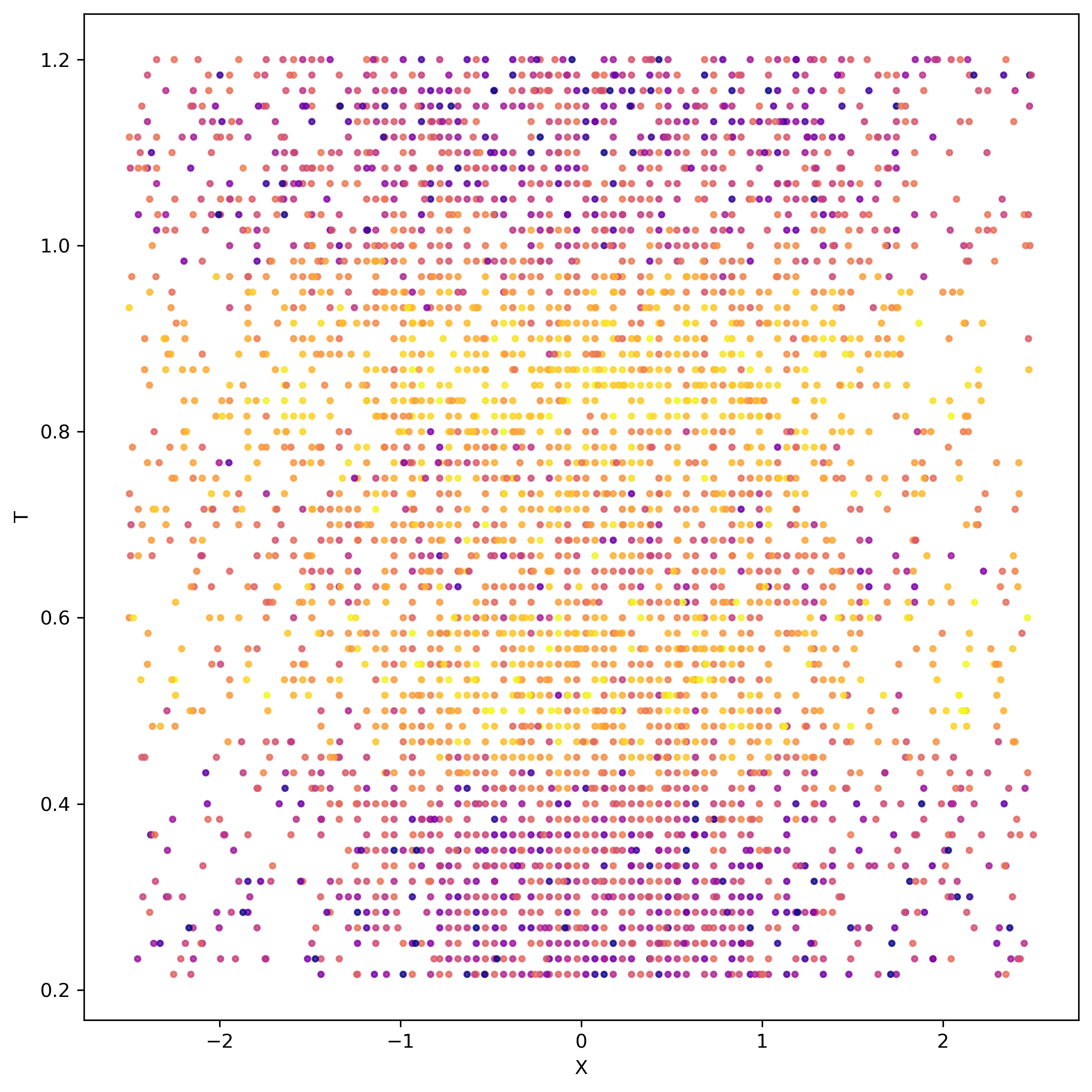}
		\end{minipage}
		\caption{Left: Reference Solution of Example 2.\label{fig:exact_sol2} Middle: DSN-PINNs Prediction.\label{fig:Eg1_pred} Right: Resampling Points.\label{fig:resample_sa2}}
	\end{figure}
	
	The numerical solution, derived from the density estimation of the SDE trajectories using the Euler–Maruyama method, serves as a reference solution \ref{fig:exact_sol2} for comparison. The spatiotemporal domain is defined as $\mathcal{Q} = [-2.5, 2.5] \times [0.2, 3.2]$ with a resampling ratio $\beta = 0.6$. The base training set $\mathcal{D}_b$ includes 240 initial, 120 boundary, and 60 spatial points, distributed across 60 time steps, yielding a total of $60 \times 60$ interior points. The mask function $f(x) = \sqrt{x}$ is used. The neural network architecture consists of 3 hidden layers, each with 20 neurons. The Adam optimizer is applied with learning rates of 0.001, 0.003, 0.005, and 0.005 for each model. In DSN-PINNs, the threshold is set to $\epsilon = 10^{-5}$, with a maximum of $N_{\max} = 6$ resampling steps and $N_{\mathrm{adaptive}} = 4$ adaptive iterations. Similarly, Figure~\ref{Eg1_error} shows the maximum absolute error distribution across models. The predicted solution using DSN-PINNs and the final resampling outcome is visualized in Figure~\ref{fig:resample_sa2}. The relevant comparison and quantitative results are also in the next section.
	
	\subsection{Example 3}
\begin{figure}[p]
		\centering
		\includegraphics[width=0.9\linewidth]{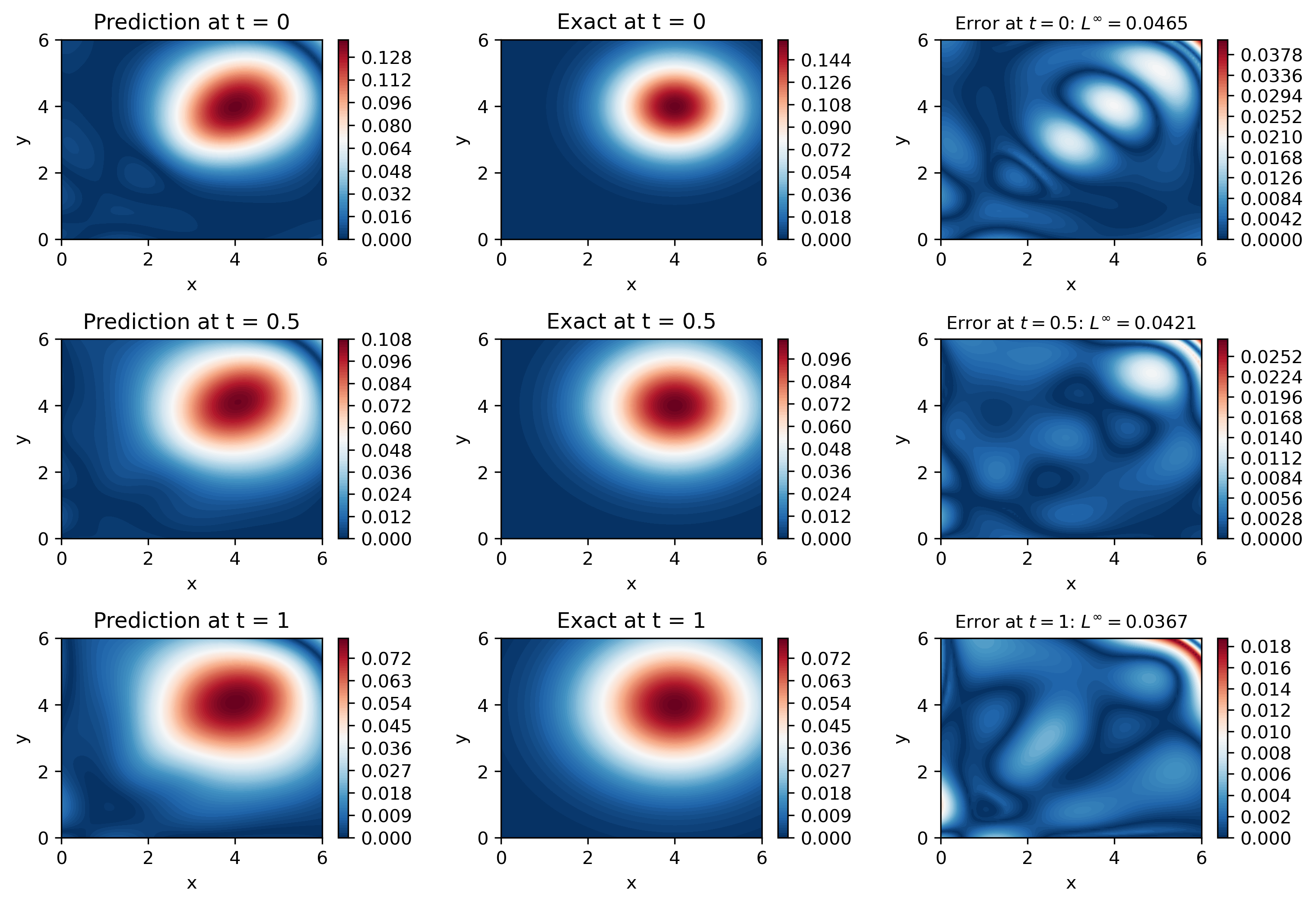}
		\caption{Standard PINNs Results at Different Snapshots. Mean PDE Residual: 0.0065. \label{BM2D_error1}}
		\includegraphics[width=0.9\linewidth]{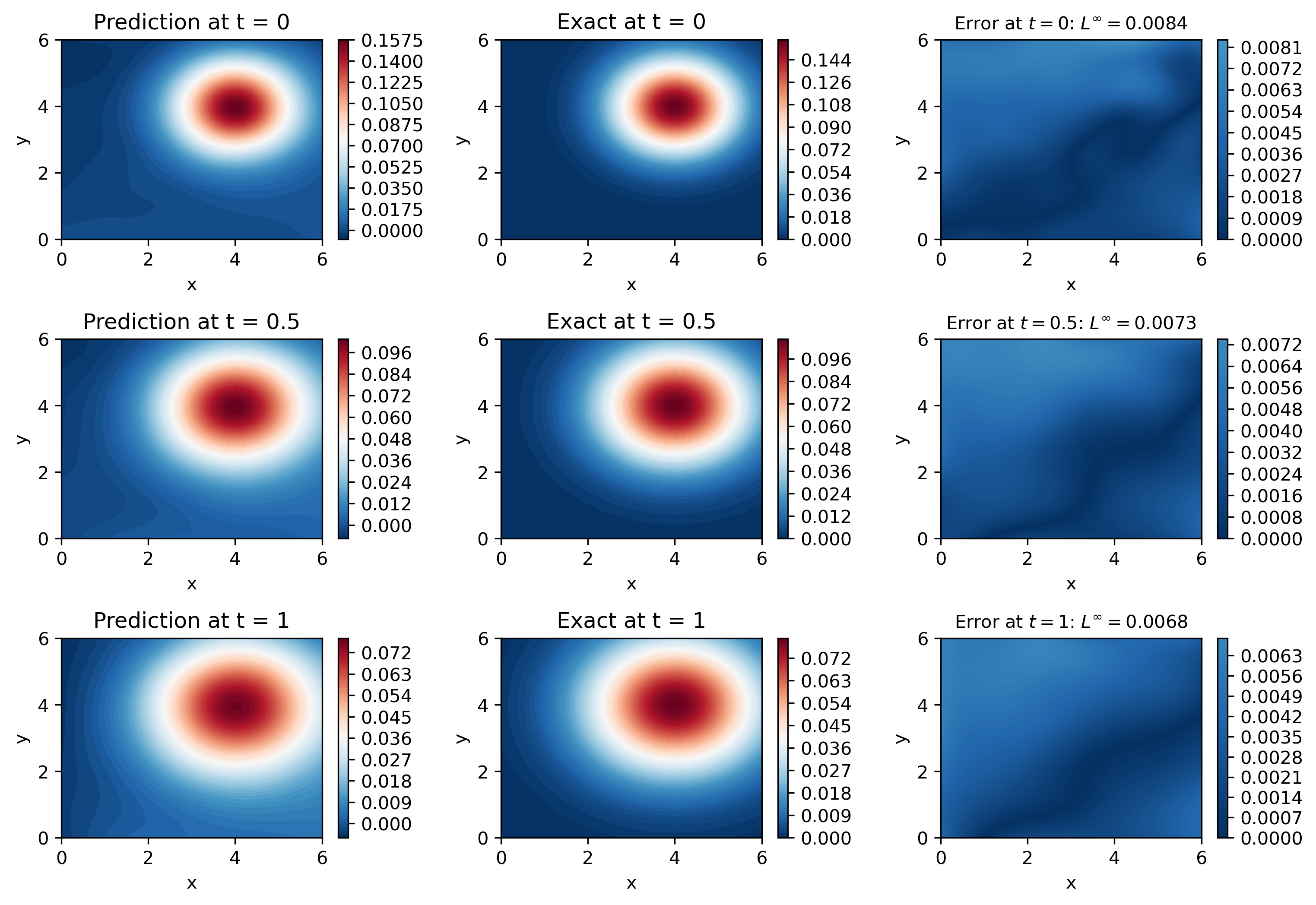}
		\caption{DSN-PINNs Results at Different Snapshots. Mean PDE Residual: 0.0014. \label{BM2D_error2}}
	\end{figure}

	We now consider some multi-dimensional examples. For a two-dimensional time-dependent FPK function as
	\begin{equation}
		\left\{
		\begin{aligned}
			&\frac{\partial p}{\partial t} - \frac{1}{2}\Delta p = 0, \\[8pt]
			& p(\boldsymbol{x}, 0) = \frac{1}{2\pi} \exp\left( -\frac{1}{2} \|\boldsymbol{x} - 4\cdot \boldsymbol{I}_2\|^2 \right).
		\end{aligned}
		\right.
	\end{equation}
	In our implementation, we use $N_{\max} = 3$, $N_{\mathrm{adaptive}} = 2$, and a threshold 
	of $\epsilon = 10^{-5}$. The spatiotemporal domain $\mathcal{Q} = [0,6] \times [0,6]$ includes 900 initial points, 640 boundary points, and 32,000 interior points generated from a $40 \times 40$ spatial grid over 20 time steps. The Adam optimizer is applied with learning rates of 0.001 for PINNs and 0.005 for DSN-PINNs. The neural network consists of four hidden layers, each with 20 neurons. A resampling ratio of $\beta = 0.7$ is used, and the mask function is chosen as $\sqrt{x}$. The absolute errors between the DSN-PINNs and the standard PINN at different snapshots are shown in Figure~\ref{BM2D_error1} and Figure~\ref{BM2D_error2}, with a detailed comparison and discussion provided in Section~\ref{sec5}.
	
	\subsection{Example 4}
	With the method’s effectiveness already established, DSN-PINNs is next applied to a 2D problem lacking an analytical solution. Consider the following 2-dimensional nonlinear oscillator as
	\begin{align}
		\mathrm{d}\begin{pmatrix} X_1(t) \\ X_2(t) \end{pmatrix} 
		&= \begin{pmatrix} X_2(t) \\ X_1(t) - 0.4X_2(t) - 0.1X^3_1(t) \end{pmatrix} \mathrm{d}t 
		+ \begin{pmatrix} 0 & 0 \\ 0 & 0.4 \end{pmatrix} \begin{pmatrix} dW_1(t) \\ dW_2(t) \end{pmatrix}\label{eq:matrix_form}.
	\end{align}
	The initial distribution is given as $ \mathcal{N}\left( \begin{bmatrix} 0 ,\ 5 \end{bmatrix}^T, \boldsymbol{I}_2 \right) $, representing a 2-dimensional normal distribution with mean $ \begin{bmatrix} 0 , \  5 \end{bmatrix}^T $ and covariance matrix $ \boldsymbol{I}_2 $. The associated FPK equation is
	\begin{equation}
		\frac{\partial p}{\partial t} = 0.2 \frac{\partial^2 p}{\partial y^2} - y \frac{\partial p}{\partial x} + 0.4 p - (x - 0.4y - 0.1x^3) \frac{\partial p}{\partial y}.
	\end{equation}
	
\begin{figure}[p]
		\centering
		\includegraphics[width=0.9\linewidth]{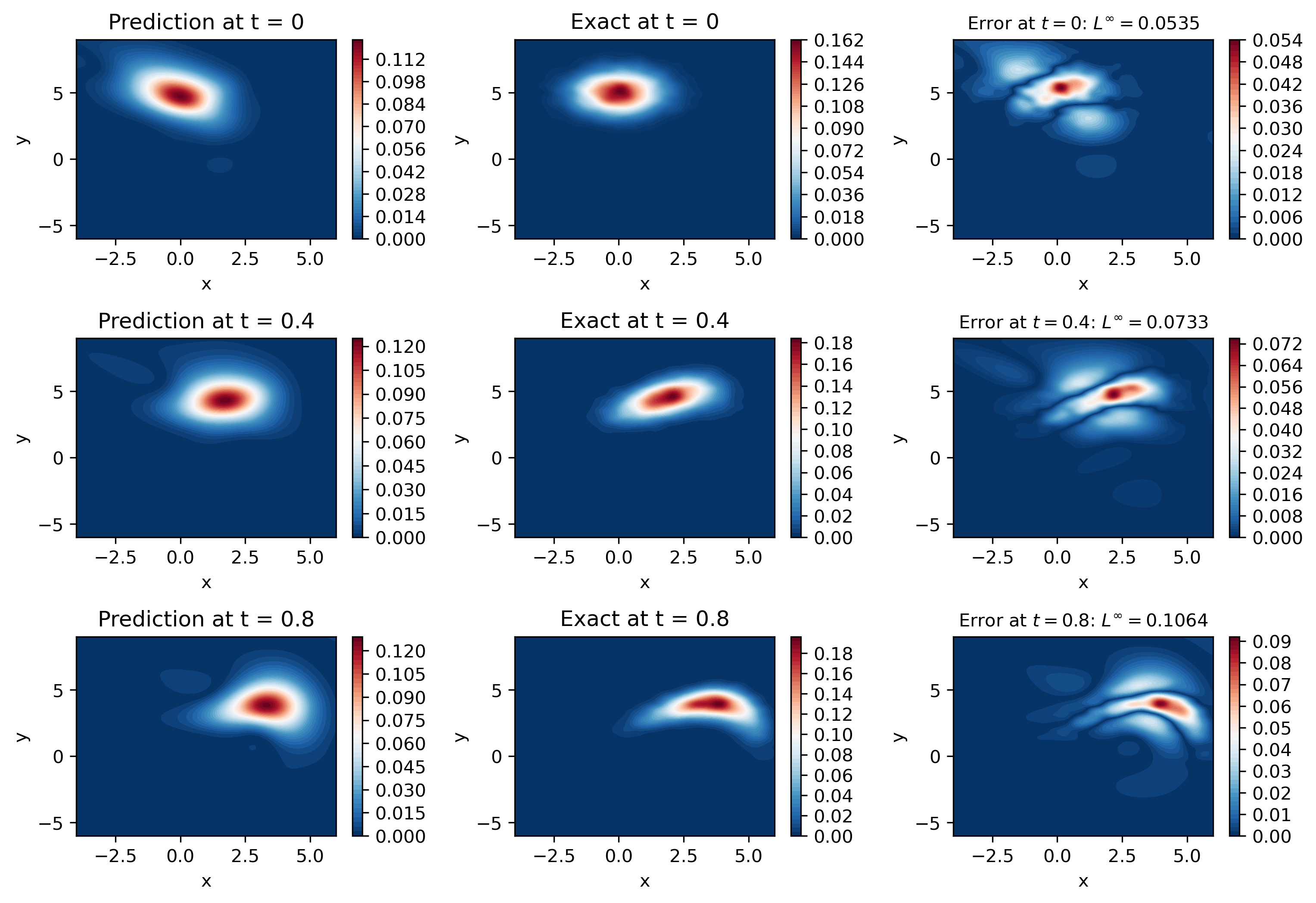}
		\caption{Standard PINNs Results at Different Snapshots. Mean PDE Residual: 0.0038. \label{Eg2_error1}}
		\includegraphics[width=0.9\linewidth]{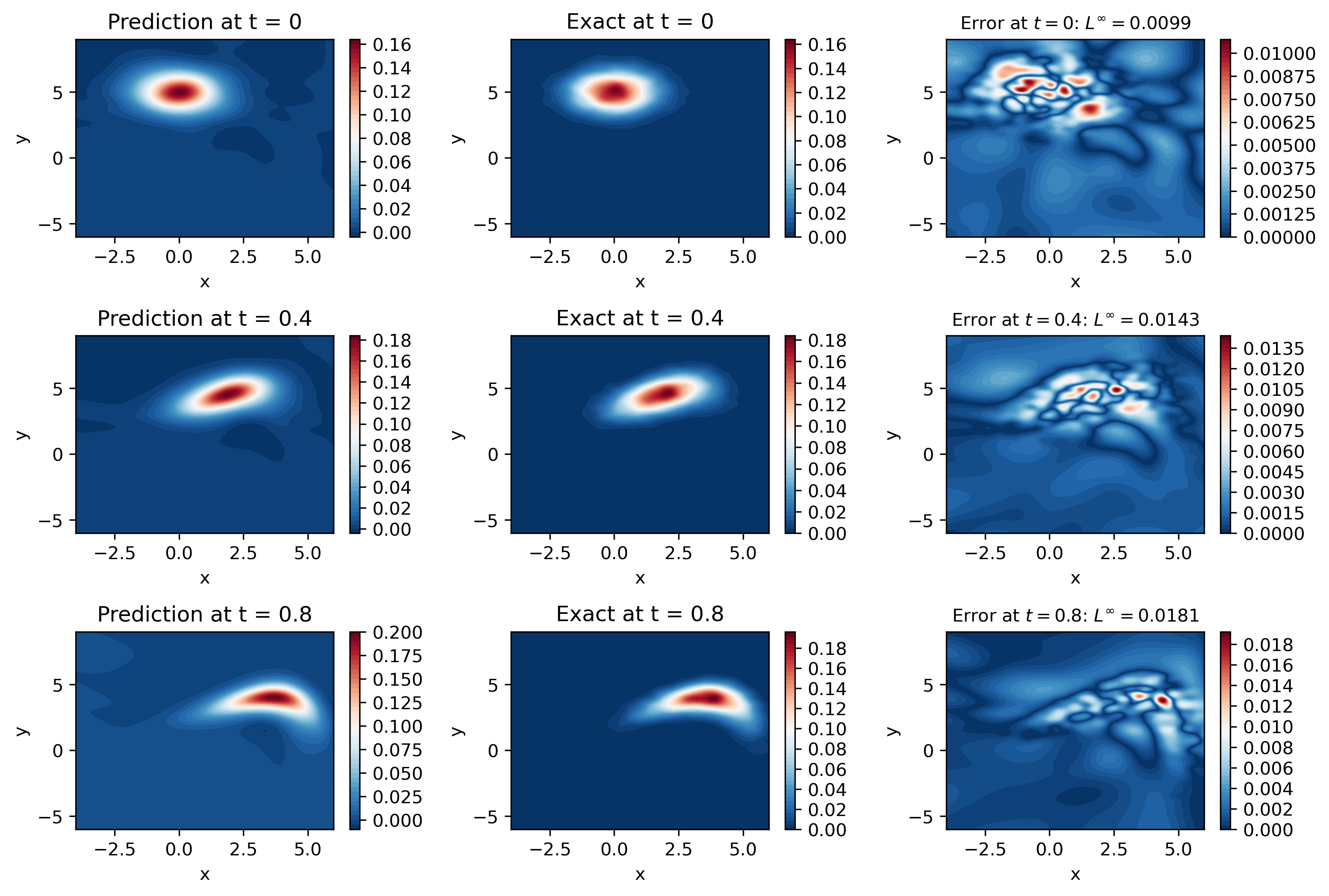}
		\caption{DSN-PINNs Results at Different Snapshots. Mean PDE Residual: 0.0018. \label{Eg2_error2}}
	\end{figure}

	In this experiment, we set $N_{\max} = 3$, $N_{\mathrm{adaptive}} = 2$, and a threshold of $\epsilon = 3\times10^{-5}$. The spatiotemporal domain $\mathcal{Q} = [-4,6] \times [-6,9]$ consists of 2500 initial points, 800 boundary points, and 32,000 interior points generated from a $40 \times 40$ spatial grid over 20 time steps. We apply learning rates of 0.005 for standard PINNs and 0.005 for DSN-PINNs. The neural network architecture includes 4 hidden layers, each containing 20 neurons. A resampling ratio of $\beta = 0.7$ is applied, and the mask function is chosen as $\sqrt{x}$. Figure~\ref{Eg2_error1} and Figure~\ref{Eg2_error2} show similar absolute error heatmaps, and their analysis is provided in Section~\ref{sec5}.
	
	\section{Result and Comparison}\label{sec5}
	
	In this section, we provide comparisons across different model variants under representative test scenarios. To evaluate and compare model performance, we adopt some standard metrics: the Maximum Absolute Error (MAE), the Mean PDE Residual $\mathcal{R}_{\mathrm{PDE}}$, and the Mean Squared Error (MSE). These metrics are defined as follows:
	\begin{itemize}
		\item \textbf{MAE}: Measures the worst-case pointwise deviation between the predicted solution $\hat{p}(\mathbf{x},t)$ and the reference solution $p(\mathbf{x},t)$,
		\begin{equation}
			\mathrm{MAE} = \|\hat{p}(\mathbf{x},t) - p(\mathbf{x},t)\|_\infty 
			= \max_{1 \le i \le N} \left| \hat{p}(\mathbf{x}^{i}, t) - p(\mathbf{x}^{i}, t) \right|.
		\end{equation}
		
		\item $\boldsymbol{\mathcal{R}_{\mathrm{PDE}}}$: Quantifies the average absolute violation of the governing PDE operator $\mathcal{N}$ evaluated at collocation points,
		\begin{equation}
			\mathcal{R}_{\mathrm{PDE}} = \frac{1}{N} \sum_{i=1}^N 
			\left| \mathcal{N}(\hat{p}(\mathbf{x}^{i}, t)) \right|.
		\end{equation}
		
		\item \textbf{MSE}: Represents the average squared difference between the prediction and the reference solution,
		\begin{equation}
			\mathrm{MSE} = \frac{1}{N} \sum_{i=1}^N 
			\left( \hat{p}(\mathbf{x}^{i}, t) - p(\mathbf{x}^{i}, t) \right)^2.
		\end{equation}
	\end{itemize}
	
	Figure~\ref{LDBM_error} and \ref{Eg1_error} display heatmaps of the absolute error between the numerical solutions and the ground truth. The error tends to decrease progressively as each model integrates further improvements. Standard PINNs only embed physical constraints into the loss function, whereas N-PINNs add the normalized condition as a soft constraint since the solutions of the FPK equation are PDFs. In addition, S-PINNs view the loss weights of training points as trainable parameters to emphasize regions with sharp transitions. Compared to N-PINNs and S-PINNs, D-PINNs adaptively resample training points in each iteration from the approximated distribution of the FPK equation, and thus yield more accurate results. DSN-PINNs combine these advantages and result in the most accurate performance among all tested models. For instance, the detailed metrics about Example 1 and Example 2 are summarized in Table \ref{tab:comparison_examples}. It follows that DSN-PINNs consistently achieve the lowest MAE, $\mathcal{R}_{\mathrm{PDE}}$, and MSE across both examples, 
	demonstrating their superior accuracy and robustness compared to all other models. 
	This confirms that combining normalization, adaptive sampling, and self-adaptive weighting effectively enhances the solution quality for the FPK equation.
	
	Next, to illustrate the representative performance gap between DSN-PINNs and the standard PINN baseline in 2D cases, $\textit{i.e.}$, Example 3 and Example 4, we compare these two models at different temporal snapshots. The error heatmap for Example 3 is shown in Figures~\ref{BM2D_error1} and~\ref{BM2D_error2}. Compared to the standard PINNs, DSN-PINNs achieve significantly lower MAE and $\mathcal{R}_{\mathrm{PDE}}$, enabling a more accurate characterization of the probability density evolution. Moreover, it effectively mitigates the accumulation of temporal errors within a certain range. Despite using a limited number of training points in the two-dimensional case, DSN-PINNs fully exploit the distributional information to deliver improved performance. Figures~\ref{Eg2_error1} and~\ref{Eg2_error2} present the corresponding results for Example 4. DSN-PINNs also provide a significantly more accurate approximation of the PDF evolution, particularly when dealing with a moderate number of training points. While standard PINNs struggle to capture the correct shape of the PDF evolution, DSN-PINNs excel in providing a more precise characterization with fewer points.
	
	\begin{table}[t]
		\centering
		\renewcommand{\arraystretch}{1.3}
		\setlength{\tabcolsep}{10pt}
		\caption{Comparison of MAE, $\mathcal{R}_{\mathrm{PDE}}$, and MSE across Different Models for Example 1 and Example 2. \label{tab:comparison_examples}}
		\begin{tabular}{@{}l l>{\centering\arraybackslash}p{3cm}>{\centering\arraybackslash}p{2.8cm}>{\centering\arraybackslash}p{2.8cm}@{}}
			\toprule
			\textbf{Experiment} & \textbf{Model} & \textbf{MAE} & \textbf{$\boldsymbol{\mathcal{R}_{\mathrm{PDE}}}$} & \textbf{MSE} \\
			\midrule
			Example 1 & Standard PINNs        & 0.0112 & 0.0042 & 2.96 $\times 10^{-6}$ \\
			& N-PINNs               & 0.0089 & 0.0029 & 3.08 $\times 10^{-6}$ \\
			& S-PINNs               & 0.0094 & 0.0023 & 1.18 $\times 10^{-6}$ \\
			& D-PINNs              & 0.0066 & 0.0010 & 1.65 $\times 10^{-6}$ \\
			& DSN-PINNs$^{\ast}$    & 0.0048 & 0.0018 & 8.10 $\times 10^{-7}$ \\
			\midrule
			Example 2 & Standard PINNs                  & 0.0720 & 0.0348 & 5.59 $\times 10^{-4}$ \\
			& N-PINNs & 0.0119 & 0.0029 & 9.86 $\times 10^{-5}$ \\
			& S-PINNs     & 0.0245 & 0.0076 & 2.26 $\times 10^{-5}$ \\
			& D-PINNs                & 0.0172 & 0.0024 & 1.63 $\times 10^{-5}$ \\
			& DSN-PINNs$^{\ast}$             & 0.0141 & 0.0017 & 1.42 $\times 10^{-6}$ \\
			\bottomrule
		\end{tabular}
	\end{table}

	\section{Experiments on Real-World Dataset}\label{sec6}
	\begin{figure}[t]
		\centering
		\includegraphics[width=1\linewidth]{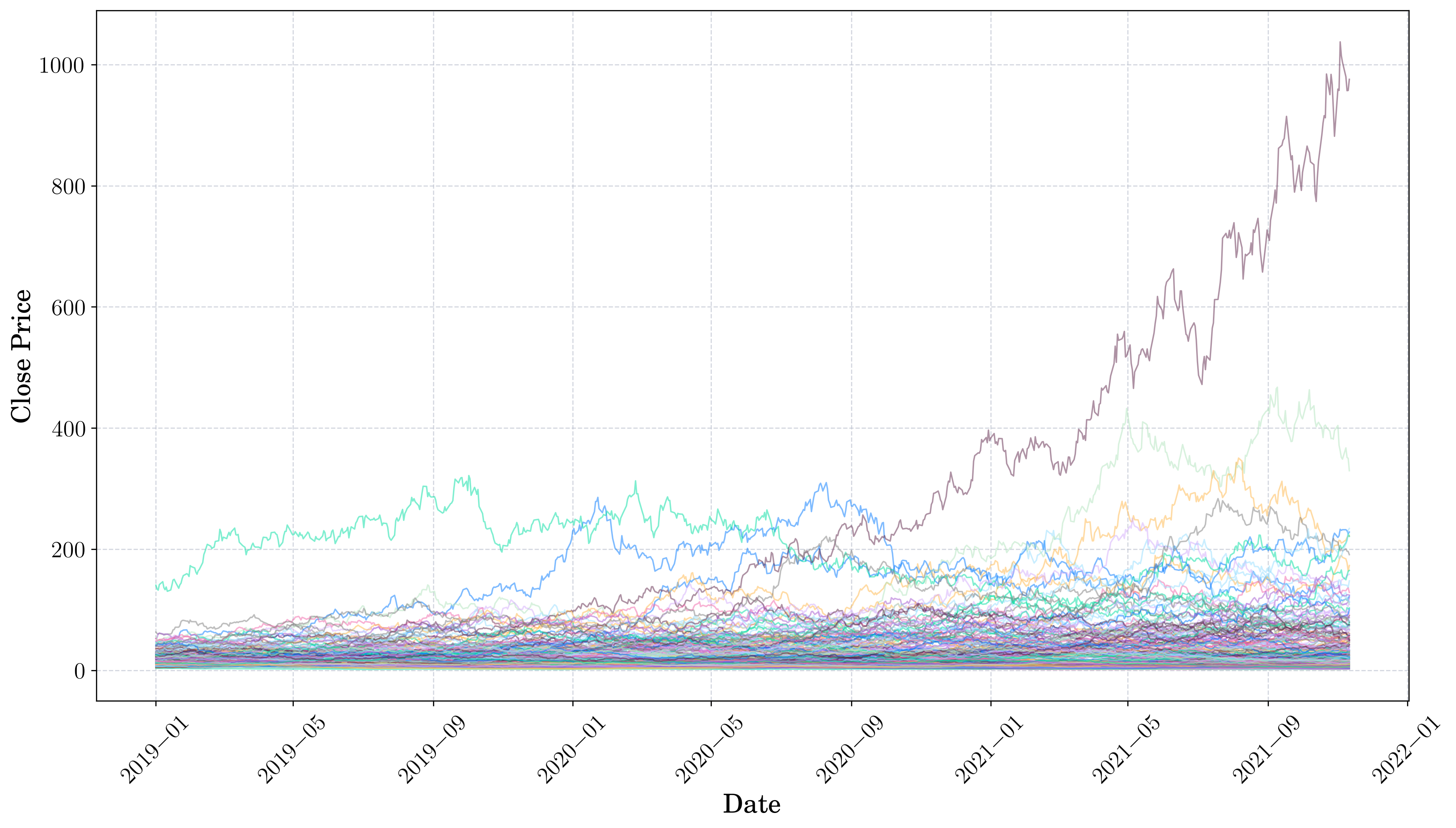}
		\caption{Sample Trajectories of Stock Prices Modeled by a Geometric Brownian Motion.}
		\label{fig:stock}
	\end{figure}
	
	As a motivating example, we collect and process daily price data of 4,500 individual stocks from the Chinese A-share market over the period 2019–2022. The stock price dynamics can be approximately described by a geometric Brownian motion (GBM) with drift \( \mu = 0.3430 \), volatility \( \sigma = 0.5693 \), and initial value \( X_0 = 13.54 \). The representative sample paths are shown in Figure~\ref{fig:stock}. The dynamics follow the SDE as
	\begin{equation}
		\mathrm{d} X_t = \mu X_t\, \mathrm{d} t + \sigma X_t\, \mathrm{d} B_t.
	\end{equation}
	
	The associated Fokker–Planck equation for the transition probability density \( p(x,t) \) is given by
	\begin{equation}
		\frac{\partial p}{\partial t} = -\frac{\partial}{\partial x}(\mu x p) + \frac{1}{2} \frac{\partial^2}{\partial x^2}(\sigma^2 x^2 p).
	\end{equation}
	
	In our setting, the initial distribution is not a Dirac delta at \( X_0 \), but a shifted log-normal density evaluated at time \( t_0 = 0.2 \)
	\begin{equation}
		p(x,0) = \frac{1}{x \sigma \sqrt{2\pi t_0}} \exp\left( -\frac{\left( \log\left( \frac{x}{X_0} \right) - \left( \mu - \frac{1}{2} \sigma^2 \right)t_0 \right)^2}{2 \sigma^2 t_0} \right), \quad t_0 = 0.2.
	\end{equation}
	The FPK equation associated with the GBM admits a closed-form solution. Given the shifted initial time \( t_0 = 0.2 \), the exact solution for the density at time \( t \) is
	\begin{equation}
		p(x,t) = \frac{1}{x \sigma \sqrt{2\pi (t + t_0)}} \exp\left( -\frac{\left( \log\left( \frac{x}{X_0} \right) - \left( \mu - \frac{1}{2} \sigma^2 \right)(t + t_0) \right)^2}{2 \sigma^2 (t + t_0)} \right), \quad x>0
	\end{equation}
	where \( X_0 = 13.54 \) and \( t_0 = 0.2 \). This corresponds to the PDF of a log-normal distribution with parameters evolving over time.
	
	\begin{figure}[p]
		\centering
		\includegraphics[width=0.84\linewidth]{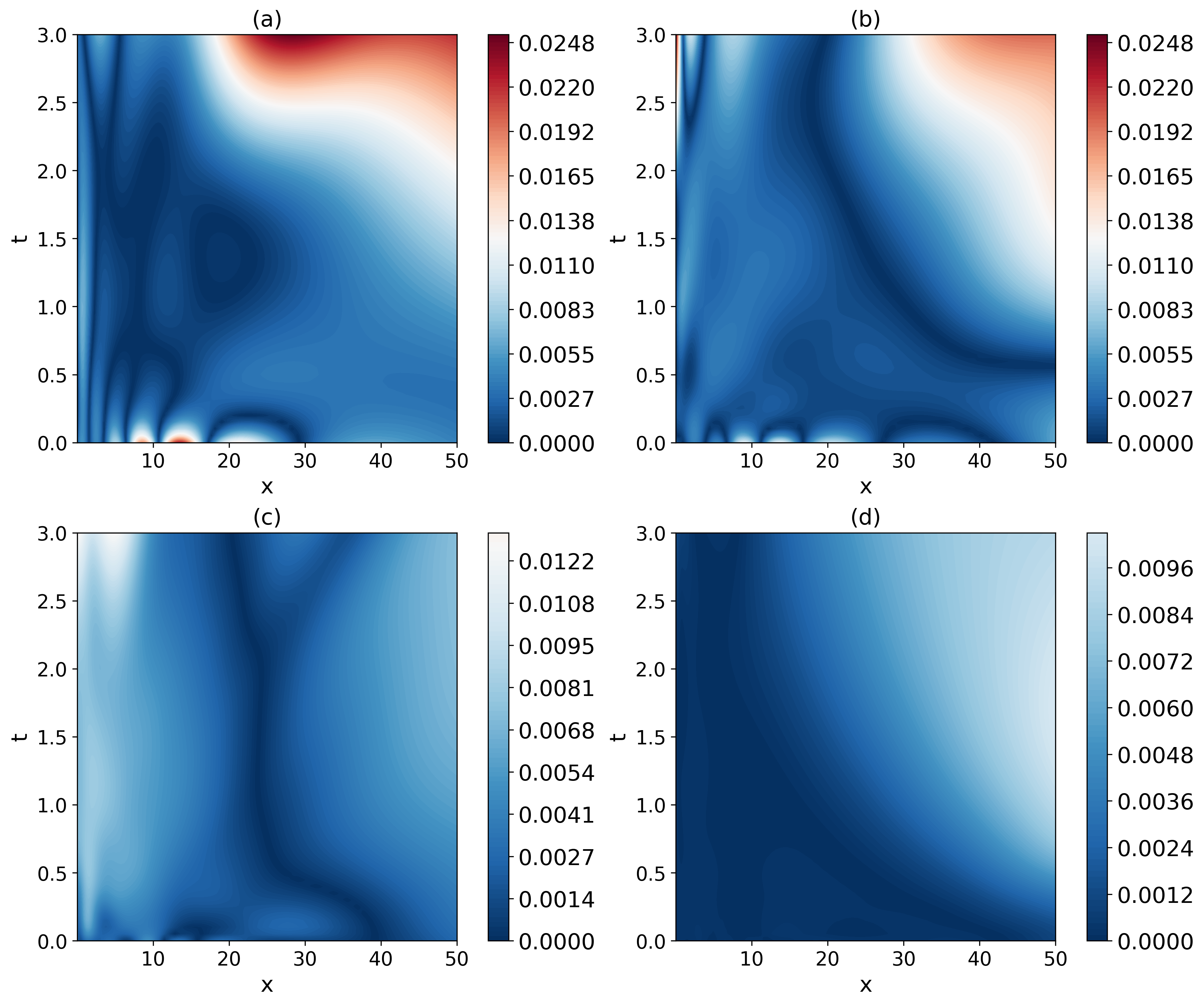}
		\caption{Absolute Errors for (a) Standard PINNs, (b) N-PINNs, (c) D-PINNs, and (d) DSN-PINNs on Real-world Data. \label{stock_error}}
		\begin{minipage}{0.49\textwidth}
			\centering
			\includegraphics[width=\linewidth]{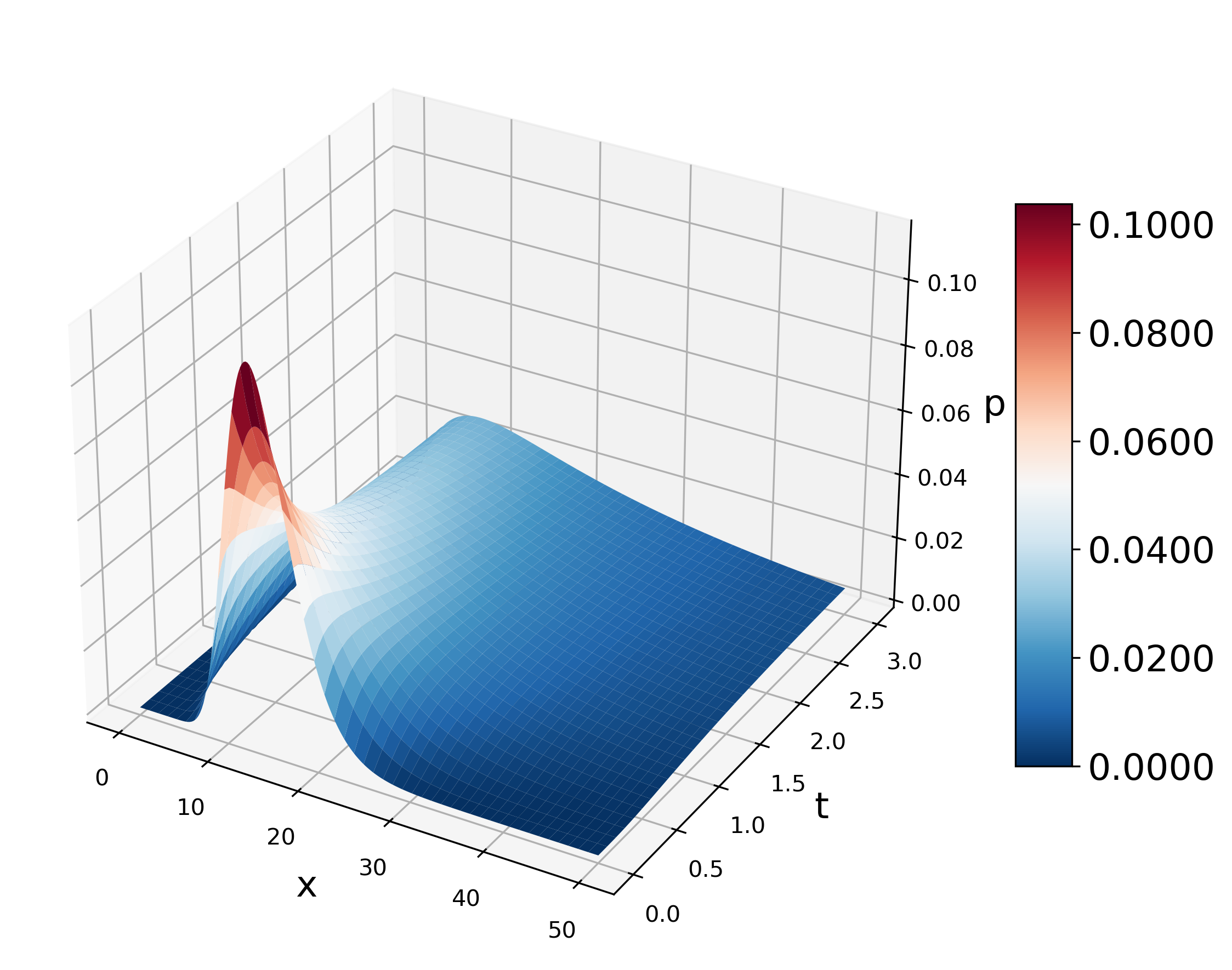}
		\end{minipage}  
		\hfill
		\begin{minipage}{0.49\textwidth}
			\centering
			\includegraphics[width=\linewidth]{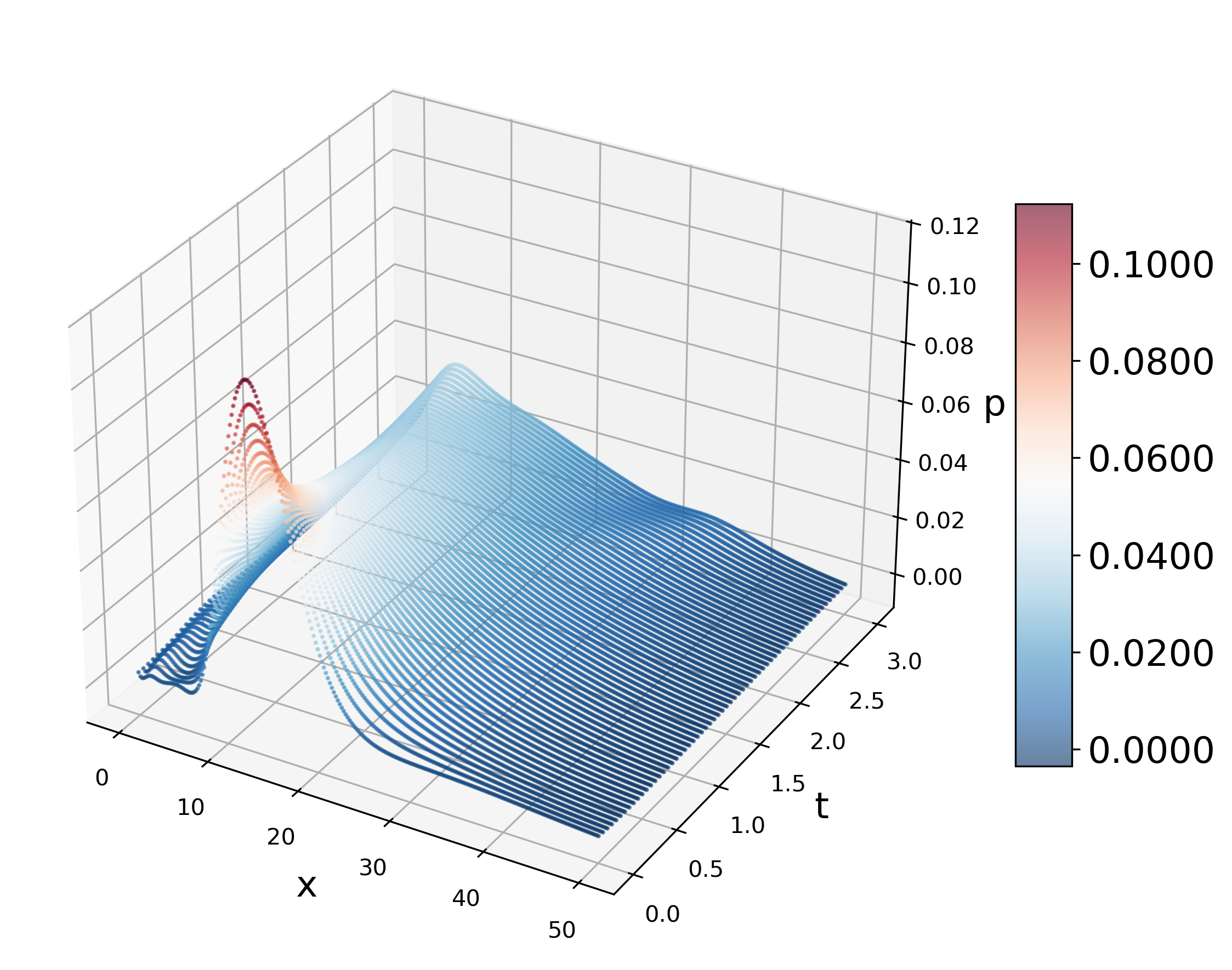}
		\end{minipage}
		\caption{Left: Reference Solution.\label{fig:exact_sol5} Right: DSN-PINNs Prediction.\label{fig:gbm_pred}}
	\end{figure}
	
	Using our DSN‐PINNs method, we process and compare this GBM problem. Except for the number of training samples and the computational domain, all settings are the same as in Example 2. We choose the spatial domain $x\in[0,40]$ (stock closing price in Chinese currency unit yuan) and the temporal domain $t\in[0,2]$, where $t=2$ represents two years of trading. Because the price range greatly exceeds the time horizon, we use finer sampling in $x$: the base training set $\mathcal{D}_b$ contains 200 initial points, 100 boundary points, and 120 interior points at each of 30 time levels, for a total of $120\times30$ interior samples. Figure~\ref{stock_error} compares the absolute errors of the standard PINNs, N-PINNs, D‐PINNs, and DSN‐PINNs. Figure~\ref{fig:exact_sol5} shows the exact Fokker–Planck solution and the DSN‐PINNs prediction, respectively. Table~\ref{tab:error_metrics5} reports the MAE, $\mathcal{R}_{\mathrm{PDE}}$, and MSE for each model. The results demonstrate that our DSN‐PINNs method retains superior accuracy and lower residuals even over the extended price interval and two‐year period. 
	
	\begin{table}[t]
		\centering
		\renewcommand{\arraystretch}{1.3}
		\setlength{\tabcolsep}{10pt}
		\caption{Comparison of MAE, $\mathcal{R}_{\mathrm{PDE}}$, and MSE across Different Models in Real-world Dataset.\label{tab:error_metrics5}}
		\begin{tabular}{@{}l>{\centering\arraybackslash}p{3.2cm}>{\centering\arraybackslash}p{3.2cm}>{\centering\arraybackslash}p{3.2cm}@{}}
			\toprule
			\textbf{Model} & \textbf{MAE}  & $\boldsymbol{\mathcal{R}_{\mathrm{PDE}}}$ & \textbf{MSE} \\
			\midrule
			Standard PINNs         & 0.0254 & 0.0040 & 5.84 $\times 10^{-5}$ \\
			N-PINNs   & 0.0252 & 0.0028 & 4.17 $\times 10^{-5}$ \\
			S-PINNs   & 0.0226 & 0.0034 &4.37 $\times 10^{-5}$ \\
			D-PINNs   & 0.0124 & 0.0007 & 2.33 $\times 10^{-5}$ \\
			DSN-PINNs$^{\ast}$   & 0.0092 & 0.0012 & 2.03 $\times 10^{-5}$ \\
			\bottomrule
		\end{tabular}
	\end{table}

	\section{Conclusion and Discussion}\label{sec7}
	In this work, we proposed the DSN-PINNs for solving time-dependent FPK equations. The method integrates a normalization-enhanced pretraining phase with a distribution-guided adaptive resampling strategy. In the pretraining stage, normalization constraints are imposed to establish a stable global structure and ensure mass conservation of the probability density, which provides a well-conditioned initialization for subsequent learning. Based on this learned prior, the training points are dynamically redistributed through weighted kernel density estimation, concentrating computational resources in regions most representative of the evolving probability distribution. This hierarchical and self-correcting mechanism enables DSN-PINNs to efficiently capture both global and local features of stochastic dynamics while maintaining computational simplicity. Extensive numerical experiments demonstrate that DSN-PINNs outperform baseline models, including standard PINNs and other relevant enhanced methods. The results reveal a consistent trend of performance enhancement with each structural refinement, normalization, and distribution self-adaptation. The DSN-PINNs method achieves the highest accuracy and robustness, showing strong capability in resolving the complex spatiotemporal evolution of probability densities and maintaining numerical stability over long time horizons. The normalization design effectively suppresses drift in the total probability, while adaptive resampling mitigates local approximation errors and sharp feature loss, leading to a balanced trade-off between accuracy and efficiency.
	
	Nevertheless, several challenges remain open for future exploration. One important direction is to extend the DSN-PINNs framework to handle FPK equations driven by Lévy noise, where nonlocal jump processes and heavy-tailed behaviors present additional numerical difficulties. Incorporating integral operators corresponding to Lévy generators and designing sampling strategies that account for discontinuities will be essential in this context. Another avenue is to integrate uncertainty quantification techniques into DSN-PINNs to systematically assess both epistemic and aleatoric uncertainties in training and predictions. Such integration could not only enhance reliability but also guide the adaptive sampling process by prioritizing high-uncertainty regions. Moreover, for high-dimensional FPK problems, particular attention must be paid to the design of training sets and normalization constraints, as approximation errors tend to accumulate with dimensionality and simulation time. Strategies such as low-rank decomposition of distributions, moment-based normalization, and annealed resampling schedules could alleviate these challenges. Additionally, exploring theoretical aspects such as convergence behavior, error propagation under iterative resampling, and the stability of the normalization constraints would provide a more rigorous understanding of the framework’s foundations. Overall, DSN-PINNs offer a promising direction for learning-based solvers of stochastic dynamical systems, and their further development toward nonlocal, high-dimensional, and uncertainty-aware formulations may substantially broaden their applicability in physics, finance, and complex systems modeling.
	
	\section*{Acknowledgement}
	This work was supported by the NSFC grant 12371198.
	
	\bibliography{references} 
\end{document}